\documentclass[12pt,flushrt,preprint,longnamesfirst]{aastex}
\usepackage{epsfig}

\newcommand{\hsi}{HS~1-029}
\newcommand{\hsii}{HS~2-104}
\newcommand{\hsiii}{HS~3-126}
\newcommand{\hsiv}{HS~4-091}
\newcommand{\hsv}{HS~5-139}
\newcommand{\hsvi}{HS~6-229}
\newcommand{\hsvii}{HS~7-289}
\newcommand{\hsviii}{HS~8-064}
\newcommand{\hsix}{HS~9-075}
\newcommand{\hsx}{HS~10-040}
\newcommand{\hsxi}{HS~11-355}
\newcommand{\hsxii}{HS~12-050}
\newcommand{\hsxiii}{HS~13-165}
\newcommand{\hsxiv}{HS~14-249}
\newcommand{\kms}{km~s$^{-1}$}
\newcommand{\ergcms}{ergs~cm$^{-2}$~s$^{-1}$}

\shorttitle{Hot Spots in SNR~1987A}
\shortauthors{Sugerman {\it et al.}}

\begin{document}
\title{Evolution and Geometry of Hot Spots in Supernova
Remnant 1987A\footnote{
Based in part on observations made with the NASA/ESA {\em Hubble Space
Telescope},
obtained from the Data Archive at the Space Telescope Science
Institute, which is operated by the Association
of Universities for Research in Astronomy, Inc., under NASA
contract NAS 5-26555.
}}

\author{Ben E.\ K.\ Sugerman\footnote{Guest observer, Cerro Tololo
 Inter-American Observatory}}
\affil{Department of Astronomy, Columbia University,
  New York, NY 10027}
\email{ben@astro.columbia.edu}

\author{Stephen S.\ Lawrence}
\affil{Department of Physics, Hofstra University,
Hempstead, NY 11549}
\email{Stephen.S.Lawrence@hofstra.edu}

\author{Arlin P.\ S.\ Crotts}
\affil{
Department of Astronomy, Columbia University,
  New York, NY 10027}
\email{arlin@astro.columbia.edu}

\and
\author{Patrice Bouchet, and Steve R.\ Heathcote}
\affil{Cerro Tololo Inter-American Observatory, Casilla 603,
  La Serena, Chile}
\email{pbouchet@noao.edu, sheathcote@noao.edu}

\slugcomment{Accepted for publication in {\em The Astrophysical Journal}}

\begin{abstract}
We present ground-based near-infrared imaging and {\em HST} optical
imaging and spectroscopy of the interaction between the ejecta of
SN~1987A and its equatorial circumstellar ring.
This interaction has made a transition, from emission
originating in just a few ``hot spots'' at restricted locations in
position angle around the ring, to a collision producing optical
emission over a nearly continuous distribution, with few breaks larger
than 45 degrees.
The centroids of the first three spots are measured
to move at $2000-3000$ \kms, which we interpret as a lower limit of
the velocity of the forward blast front.
Multi-wavelength light curves of the spots show that they do not
evolve uniformly, and change significantly on timescales as short as
one month; in particular the first spot shows a significant break in
its lightcurve. 
%, and may be approaching a plateau in intensity.
%
Implications of observed delays between spots appearances are
discussed, which leads to a generalized model of hot
spot evolution, and suggests that the early appearance of the first
hot spot is explained by its inward radial position and a fairly
uniform forward blast wave, rather than extraordinary physical
circumstances.
Data further suggest that the forward blast is reaching the bulk of the
inner ring material to the east, the density of which appears 
higher than elsewhere in the ring.
We study the ring geometry, finding evidence suggestive of an intrinsic
ellipticity of 0.95, and find lower and upper distance limits of 
$47.9\pm0.92$ kpc and $54.4\pm2.1$ kpc, respectively.
\end{abstract}


%%%%%%%%%%%%%%%%%%%%%%%%%%%%%%%%%%%%%%%%%%%%%%%%%%%%%%%%%%%%%%%%%%

\section{INTRODUCTION \label{sec-1}}

Supernova (SN) 1987A in the Large Magellanic Cloud is the first
naked-eye SN in over three centuries, and the first SN remnant (SNR)
seen to form within a pre-existing circumstellar environment that has
been mapped in significant detail.
It serves as a vital test bed for the colliding winds model of
mass-loss nebulae and the interaction of SN ejecta with
interstellar and circumstellar material (CSM); an important rung in the
cosmological distance ladder;  a valuable probe of the interstellar
medium (ISM);  
and a unique laboratory for studying SNR formation and the final
stages of massive star evolution.

Narrow emission lines in {\em International Ultraviolet Explorer}
({\em IUE}) spectra at day 80 (after SN) indicated the existence
of a circumstellar nebula \citep{Fra89} which was confirmed
in long-slit [\ion{O}{3}] and H$\alpha$ spectroscopy on day 300 by
\citet{WR89}.  The first resolved images were taken in
these same bands by \citet{CKM89} on day 750, and
by \citet{Wam90} on day 1037, and indicated the existence of an
elliptical structure with a central cavity, the latter evidenced by the
rapid fall of radio signal from ejecta--CSM
interaction shortly after the SN was discovered \citep{Tur87}.
{\em Hubble Space Telescope} ({\em HST}) imaging
\citep{Jak91,Jak93,Jak94,Pla95,Bur95}
revealed this structure to be a dense circumstellar equatorial ring (ER)
flanked by two larger outer rings, while kinematics of the
ER proved it to be a planar ring, expanding at $\sim10.3$ \kms\ and
inclined at $\sim43\arcdeg$, rather than a limb-brightened ellipsoid
\citep{CH91}.

In his model of self-similar interaction of ejecta with ambient media,
\citet{Che82} demonstrated the development of a double-shock system,
in which a forward blast wave is driven into the CSM and
a reverse shock is driven into the expanding ejecta, with a contact
discontinuity between the shocked CSM and shocked ejecta.
Models predicted this forward shock would impact the ER
between 13--20 years after the SN explosion
\citep{Luo91,Luo94,CD95,BBM97b}.
Making simple assumptions of symmetry, this impact was expected to
appear as an initial brightening at the northern (closest) segment of
the ER, and then spread in P.A.\ to engulf the entire ring
within roughly 1 year, due to light-travel delays across the ER.

The first indications that the double-shock scenario was correct came in
1997 when a 250 \kms\ blue-shifted H$\alpha$ feature was observed at
P.A.\ 29\degr\ \citep{Pun97} and a Ly$\alpha$ feature was observed
with velocity $\lesssim1.5\times10^{4}$ \kms\
\citep{Gar97b} using the Space Telescope Imaging
Spectrograph (STIS) aboard {\em HST}.
This H$\alpha$ feature was quickly  confirmed in Wide Field and
Planetary Camera 2 (WFPC2) imaging \citep{Gar97a}, and was interpreted
as the first ``hot spot'' from the impact of high-velocity
forward shock with the ER \citep{Son98}.  The Ly$\alpha$ feature has
been modeled and interpreted as neutral hydrogen in the CSM
between the SN and the ER crossing the reverse shock
\citep{Son98,Mic98,Mic99,Mic00a}.

Contrary to the simplest expectations, no new loci of ER-ejecta
interaction were discovered following the first hot spot for nearly
three years, until \citet{Bou00} reported a new brightening near
P.A.\ 104\degr\ in ground-based \ion{He}{1} imaging, which was soon
followed by the discovery and confirmation of five additional spots in
{\em HST} data \citep[hereafter Paper I]{Law00,Mar00,Gar00,LSC00,LSB00}.
In Paper I, we re-analyzed the WFPC2 data using
PSF-matched difference imaging, and traced the appearance of the first
hot spot (\hsi\ using Paper I notation\footnote{The first number
represents the order of discovery, and the trailing three digits
encode the first reported imaging P.A.\ of the feature
}) to 1995 March, as well as most of
the newly-discovered spots to early 1999.

We are now observing a unique period in the formation of SNR~1987A as the
high-velocity SN debris overtakes the slowly expanding ER.
While this interaction will eventually destroy the circumstellar
nebula, it will provide unique opportunities in the coming years to
explore the complicated products of stellar mass loss, by detailing the
circumstellar environment directly through shock-heating.
Furthermore, the X-ray and UV flux from this interaction will reionize
unseen portions of the nebular structure, revealing many of them for
the first time.

The visible consequences of the forward shock's impact on the ER
are apparent and developing rapidly.
In \S\ref{sec-obs} we report new ground-based and STIS observations
taken of SNR 1987A.  Results are presented in \S\ref{sec-res}.
We show that between the years 2000 and 2001, the number of confirmed
hot spots has nearly
doubled (\S\ref{ssec-disc}), with spots now located in a nearly
continuous distribution about the ER (\S\ref{ssec-loc}).  We also show
in \S\ref{ssec-loc} that the centroids of the first three hot spots
display proper motion.  We study the geometry of the ER in
\S\ref{ssec-param}, which we use to build a
deprojected view of the system in \S\ref{ssec-deproj}.
Light curves for the hot spots are discussed in
\S\ref{ssec-evol}, and the bulk interaction between the ejecta and ER
is examined in \S\ref{ssec-bulk}.
In \S\ref{sec-tests}, we examine the accuracy of our detection and
photometry methods.  We discuss a general model for hot spot evolution
in \S\ref{ssec-gevol}, and consider the implications of hot spot
timing in \S\ref{ssec-time}, followed in \S\ref{ssec-dist} by a brief
discussion regarding the distance to the SN.

%%%%%%%%%%%%%%%%%%%%%%%%%%%%%%%%%%%%%%%%%%%%%%%%%%%%%%%%%%%%%%%%%%

\section{OBSERVATIONS AND REDUCTIONS \label{sec-obs}}

We analyze data taken on the Cerro Tololo Inter-American
Observatory (CTIO) 4-m telescope with tip-tilt first-order wavefront
correction, as part of an ongoing ground-based monitoring campaign of
the SN and its circumstellar environment.  Data were taken on 2001
March 17--18 [3.7~hr total integration] and 2001 November 27--28
[1.5~hr total integration] with the OSIRIS imager
in the \ion{He}{1}~1.084$\mu$m line, and
differenced from data reported in Paper I
(1998 October 6 from CIRIM, 1.5~hr total integration; 1999 December 25
from OSIRIS, 3.5~hr total integration)
using {\em difimphot}.

\begin{figure}
\centering
\epsscale{1.}
%\plotone{../pldate.ps}
\plotone{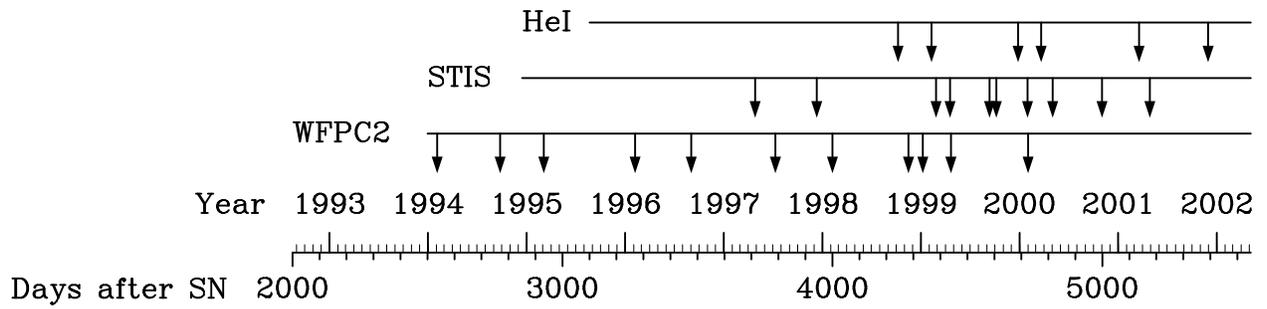}
\caption{Schematic relating calendric dates to the number of days
after the SN exploded, as well as showing the dates of the WFPC2,
STIS and ground-based \ion{He}{1} observations used in this paper.
\label{pldate}}
\end{figure}

We analyze public and Director's Discretionary data from the {\em HST}
archive, making use of:
\begin{itemize}
\item STIS spectra (G750M grating, 2\arcsec slit)
from 1997 April 26 [day 3714], 2000 May 1 [day 4816], 2000 October 21
[day 4996] and 2001 April 27 [day 5176];
\item STIS F28X50LP imaging taken between 1997 December and 2001 April
[days 3941--5176]; and
\item WFPC2 images taken through
F336W, F439W, F555W, F656N, F658N, F675W, and F814W between
1994 February and 2000 February [days 2537--4725].
\end{itemize}
For clarity, Figure \ref{pldate} shows the correspondence between
the number of days following the SN, the calendric date, and
the epochs of all data mentioned above.

Pipeline reduced {\em HST} spectra and imaging were taken
directly from the archive, and aberrant pixels (listed in the data quality
and hot pixel files\footnote{available at
{http://www.stsci.edu/instruments/wfpc2/wfpc2\_hotpix.html}}) were
fixed using the method to be described shortly.
STIS spectra were co-added using cosmic-ray
rejection, and finally wavelength calibrated using {\em
stsdas} within the {\em IRAF} data reduction and analysis system\footnote{
{\em IRAF} is distributed by the National Optical Astronomy Observatories,
which are operated by the Association of Universities for Research
in Astronomy, Inc., under cooperative agreement with the National
Science Foundation.
}.
Cleaned WFPC2 and STIS images were summed with cosmic-ray
rejection, and Stars 2 and 3 were removed using {\em Tiny Tim} model
point-spread functions (PSFs), and geometric
distortions in the PC chip \citep{Hol95} were corrected using
the {\em drizzle} routine with a 1:1 resampling.  Images were finally
unsharp masked to remove any sky background.
As in Paper I, we apply the {\em difimphot} image subtraction techniques of
\citet{TC96}, which employ Fourier techniques to match
empirically-derived stellar profiles between two images.  This
technique is limited to data with resolution better than $\sim2.2$
pixels per FWHM, an effective critical minimum required for Fourier transforms
of imaging data \citetext{R.~Uglesich, private communication}.  Since
the  PC and STIS images
have a nominal resolution of $\sim1.7$ pixels, we
convolve the data with a circular Gaussian of
$\sigma=0.65$ to achieve a final stellar PSF of ${\rm FWHM}\gtrsim2.3$
pixels.  Data are finally geometrically registered to a common
orientation with residuals $\le 0.05$
pixel rms (PC) and $\le 0.1$ pixel rms (STIS).  Aperture and
crowded-field photometry was performed using standard {\em daophot}
\citep{Ste87} techniques with {\em Tiny Tim} model PSFs processed by
identical procedures as the data to mimic the registration steps.
PC fluxes were calculated using {\em synphot}, including decontamination
corrections, while STIS imaging fluxes are presented in instrumental counts
 sec$^{-1}$.

As briefly outlined in Paper I, hot pixels account for
$\sim1\%$ of both the PC and STIS chips, and since most images were
not spatially dithered, these pixels present a stochastic and
non-Gaussian source of noise.  During the reduction procedures
(registration and smoothing), single hot pixels can take on
point-source-like profiles, thereby creating a strong source of
false signals.  We correct known bad pixels by replacing each
with the median value of the 8 surrounding pixels, taken from an
initial cosmic-ray cleaned image.  The difference between the original
and cleaned images is a ``contamination image,'' i.e.\ a map of the bad
pixels locations, which accompanies each data image through each step
of the reduction pipeline.  Since many data sets consist of only two
images, often not spatially dithered, we recognize that this
correction technique is predominantly cosmetic, and hence
to minimize false detections in our analysis, a potential hot spot signal
must not lie within a bad-pixel domain, and must appear in at least
two difference images created from mutually exclusive epochs.

%%%%%%%%%%%%%%%%%%%%%%%%%%%%%%%%%%%%%%%%%%%%%%%%%%%%%%%%%%%%%%%%%%
\section{RESULTS \label{sec-res}}

\subsection{Discovery of New Hot Spot Activity \label{ssec-disc}}

\begin{figure}
\centering
\epsscale{1.}
%\plotone{../HeI.ps}
\plotone{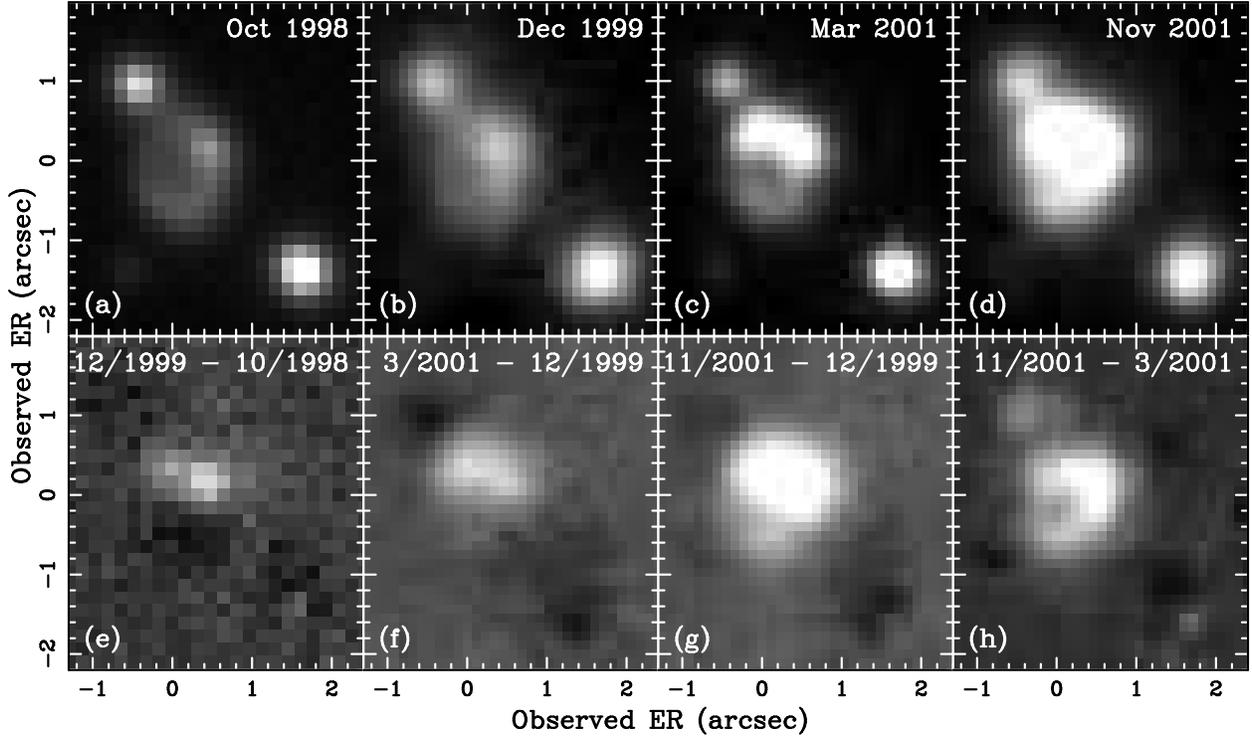}
\caption{Ground based CTIO 4 m \ion{He}{1} 1.083$\mu$m imaging of
SNR~1987A.  North is to the right and east up.  (a) CIRIM data from 1998
October. (b) OSIRIS data from 1999 December. (c) OSIRIS data from 2001
March. (d) OSIRIS data from 2001 November.
(e) Difference image between 1999 December and 1998 October.  We see
the brightening of \hsi, the appearance of spots to the southeast,
and the fading ER to the west.
(f) Difference image between 2001
March and 1999 December.  Flux from spots 1--5 continues to increase, and
unresolved spots distributed about the ER are now brightening
faster than the fading ER material.
A small residual of star 2 (lower right) remains since
this star's flux is within the non-linear regime of the detector in
the 1999 image.
(g) Difference image between 2001 November and 1999 December.  Flux
has dramatically increased along the eastern half of the ER, and the
western half is now standing out against the background, suggesting a
population of brightening spots.
(h)  Difference image between 2001 November and 2001 March.  This
image highlights the smaller-scale changes in flux: spots appear to
now be located around the entire ER, except in the south-southwest
region.  A population of spots surrounding \hsi\ may explain the
``L''-shaped flux distribution in the northeast quadrant.
Star 3, at the upper left, is known to be variable, and as seen in the 
difference-image sequence, was brighter in 1999 December and 2001 November 
than in 2001 March.
 \label{HeI}}
\end{figure}

As in Paper I, we perform difference-image analysis to search for new
hot spots.  The contrast between hot spots and the
underlying ER is greater in \ion{He}{1}~1.083$\mu$m than in optical
lines at early times \citetext{Paper I}, thus \ion{He}{1} imaging is a
sensitive probe of
new ER-ejecta interaction.  Figure \ref{HeI} displays a tiling of
images from our ground-based monitoring. Panel (a) shows the ER and
\hsi\ resolved between Stars 2 and 3 on 1998 October 6, taken with
the CIRIM imager.  Panels (b)--(d) show data from the OSIRIS
imager taken on 1999 December 25; 2001 March 17--18; and
2001 November 27--28 (as noted in each window).
%Panels (a)--(d) are displayed
%such that Star 2 (lower right-hand corner) has the same brightness.
Panel (c) clearly shows the brightening of the spot 2--5
complex, and the fading of Star 3 (a known variable star). In panel
(d), we see that hot spot flux has dramatically brightened in eight
months, nearly engulfing the entire ER.

Panels (e)--(h) display PSF-matched difference images made with the
above data, as noted in the top of each window.  The discovery
data from \citet{Bou00} is shown in panel (e).  This difference image,
between 1998 and 1999, was photometrically
scaled using the fluxes of stars of constant brightness located
throughout the field.  As such, white regions have brightened
and black regions have faded since the earlier epoch.  We see the
continued brightening of
\hsi, the faint appearance of \hsii, and the fading ER to the west.
Panel (f) displays the data from 2001
March subtracted by 1999 December, again scaled photometrically to
constant stellar flux.  While hot spot emission from P.A.\
30\degr--150\degr\ is clear, comparison of the western half of the ER to
that in panel (e) reveals that the fading ER has been offset
by a continuous locus of brightening flux.  This trend is confirmed in
panel (g), a difference image made between 2001 November and 1999
December, in which we now see the southwestern limb of the ER
standing out distinctly from the background.  To better examine the
extent of this brightening, we display the difference between  2001
November and 2001 
March in panel (h).  Flux is evident along the entire ER except
the region roughly located between P.A. 160\degr\ and P.A. 210\degr;
even this latter region must be brightening marginally, otherwise the
fading of the underlying ER would result in negative (black) pixels.
At this resolution, we can not be certain of the actual distribution
of sources producing this continuous, positive image superposed on the
fading ER material.  We consider it realistic that this
reflects unresolved emission from a population of  young hot spots
distributed roughly uniformly about the ER.  Flux has increased around
the position of \hsi\ the most, yet the flux distribution at
this position is elongated (south and westward), notably differing
from the image's PSF.  This
is suggestive that the brightening is not due to \hsi\ alone, but can
be attributed to the rapid brightening of a population of spots on
either side of \hsi.

\placefigure{HeI}

While the above ground-based data are effective at describing the global
behavior of the ER-ejecta interaction, we study the smaller-scale
evolution with {\em HST}.
The ER is resolved well enough to allow us to compensate for its
fading by scaling images to be differenced by
its flux (rather than by constant stellar photometry),
which we measure in elliptical arcs in both images.  For most
epochs, we employ an arc of radial width $0\farcs6$
between P.A.\ 300\degr--10\degr, however as more of
the ER becomes contaminated with new spots, we co-add several smaller
arcs distributed about the ring.  Since the ER fades non-uniformly,
the ring is often over-subtracted in some places, and stellar
residuals often remain.

\begin{figure}
\centering
\epsscale{1.}
%\plotone{../images.ps}
\plotone{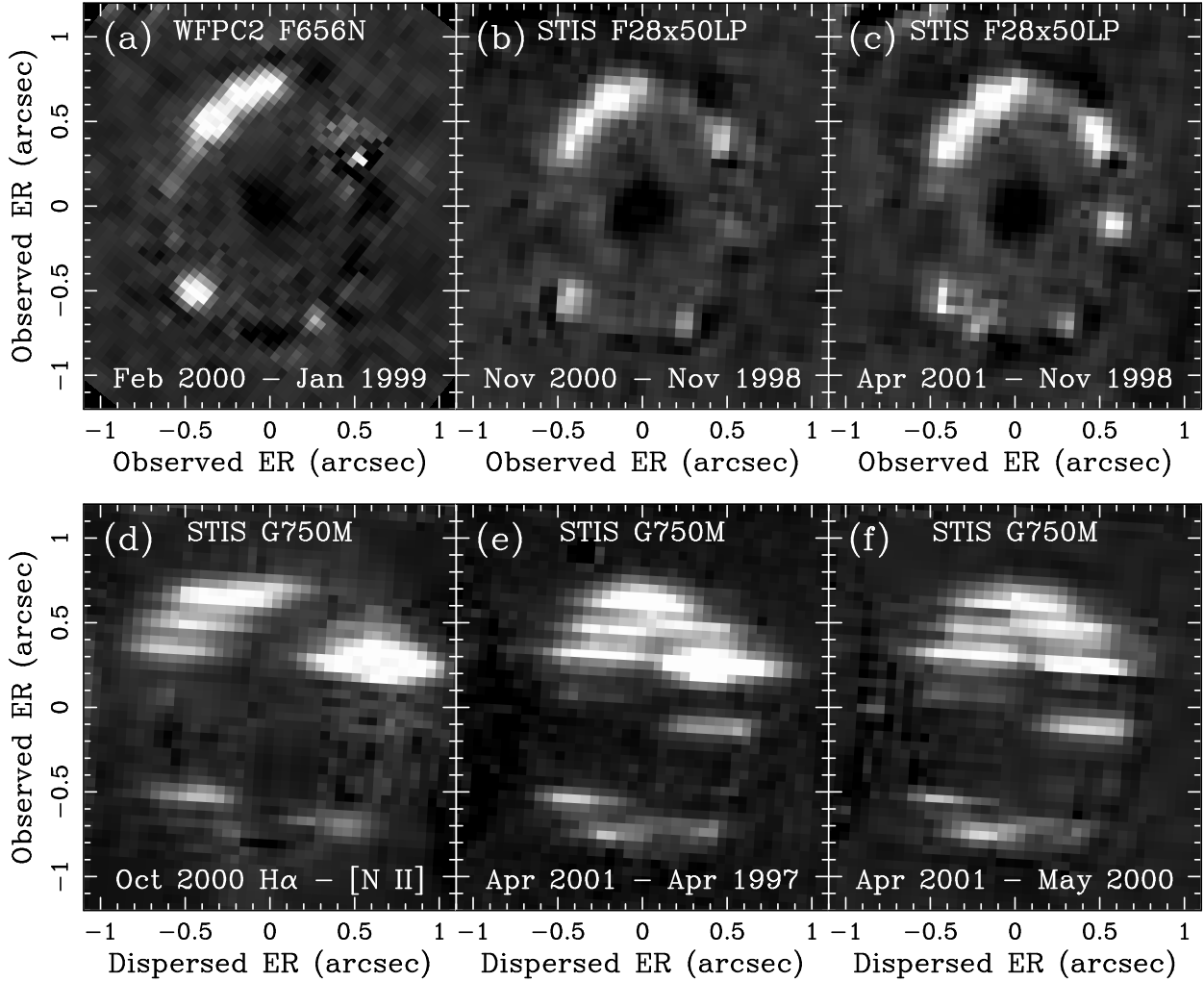}
\caption{{\em HST} difference imaging of SNR~1987A. North is right and
east is up.  (a) WFPC2 F656N difference image between 2000 February and
1999 January.  (b) STIS F28X50LP difference image between 2000
November and 1998 November. (c) STIS F28X50LP difference image between
2001 April and 1998 November.  \hsi\ has been removed from images in
panels (a)-(c) using a {\em Tiny Tim} model PSF.
(d) STIS G750M H$\alpha$ spectral image from 2000 October with the
[\ion{N}{2}] $\lambda6583$ image subtracted away to remove ER flux.  In
this orientation, wavelength increases to the right. (e)--(f)
STIS G750M H$\alpha$ spectral image from 2001 April differenced from
identically-oriented H$\alpha$ observations in 1997 April and 2000 May
(respectively), to remove ER flux.  In this orientation, wavelength
increases to the left, and one pixel equals 0.56 \AA\ or $\sim 50$ \kms\ of
Doppler shift.
 \label{images}}
\end{figure}

\placefigure{images}

Figure \ref{images}a shows a difference image in F656N\footnote{At the
systemic recessional velocity of SN~1987A ($-289$ \kms), the
F656N filter transmits in vacuo more efficiently  [\ion{N}{2}]
$\lambda$6548 than H$\alpha$, using pre-launch filter curves.  Since
no concurrent images in F656N and F658N have been taken, it is not
straightforward to measure a monochromatic flux in H$\alpha$ through
this filter.}
between 2000 February 2 and 1998 February 5.
The flux from \hsi\ has been removed using the {\em Tiny Tim} model PSF,
however a small core remains.
We clearly see the first six spots, as reported from these data by
\citet{Gar00} using independent reduction techniques, as well as \hsvii\
reported by \citet{LSC00}. A faint
signature of \hsx, as well as marginal features at P.A.\ 174\degr\
and 314\degr, as reported by \citet{SLC00} from 2000
May 1 STIS
observations, are also detectable.  Since stellar residuals remain due
to the ER scaling, flux from \hsvi\ is a combination of emission from
the hot spot and coincident Star 5 (P.A.=230\degr, $r=0\farcs73$).

In light of the rapid developments in hot spot discovery in early
2000, we took STIS spectra and imaging on 2000 May 1, 2000 October 29,
and 2001 April 27
(through Director's Discretionary observing time on {\em HST}).
Figure \ref{images}b shows a difference image made in the F28X50LP
imaging filter, between 2000 November 3 and 1998 November 14.
Again, flux from \hsi\ has been removed.  In addition to the
aforementioned spots, \hsxi\ \citep{SLC00} can be seen.  Figure
\ref{images}c shows a similar difference image in F28X50LP between
2001 April 27 and 1998 November 14.  A new feature slightly eastward
of \hsx\ is apparent once \hsi\ is removed, and faint signatures from P.A.s
163\degr, 248\degr\ and 314\degr\ also appear.

To create a systematic observing strategy which maximizes the
information gathered in a single spectrum, while dovetailing with
previous observations with STIS,
we imaged the entire ER in the G750M grating with the
same $2\arcsec$ slit orientation as the 1997 April spectrum
for the 2000 May and 2001 April observations, and rotated by 180\degr\
for 2000 October. This produces
spatially-resolved monochromatic images of the entire ER in a large
number of transitions with a velocity scale of $\sim 25$ \kms\ pix$^{-1}$.
Since the ER has a rest-frame expansion velocity of 10.3 \kms\
\citep{CH91}, even nascent hot spots are clearly detectable
by the Doppler shifting of their high-velocity emission away from the ER.
Furthermore, the identical slit orientations allow direct
spectral image subtraction from previous epochs, preserving data
quality while easily removing the background ER flux.
Results reported from Paper I
were generated from a direct subtraction of the
1997 and 2000 May spectra, however since the 2000 October spectrum has
no identically-oriented antecedent, we scaled the neighboring
[\ion{N}{2}] $\lambda$6583 image to remove the H$\alpha$ ER image.
The flux of \hsi\ was 15 times greater in H$\alpha$ than
[\ion{N}{2}] in a $0\farcs1$ STIS spectrum from 1999 August 30, hence
this ER removal should not affect the discovery of new spots. 
In Figure \ref{images}d, we see an asymmetric tail toward the east
from \hsi, identified by \citet{SLC00} as \hsx, as well as a faint
smear at the location of \hsxi.  Furthermore, marginal detections are
seen at approximate P.A.s 170\degr,
250\degr, and 315\degr.  Figures \ref{images}e--f show direct spectral
subtractions of the 2001 April H$\alpha$ image from that taken in 1997
April and 2000 May, respectively.  Since the slit is oppositely
oriented from that in panel (d),
the direction of Doppler shifting is reversed.  \hsxi\ is unambiguous,
as are spectral smears around P.A.\ 50\degr\ and 250\degr; a faint
signature is detected at P.A.\ 165\degr\ and we again note the marginal
flux around P.A.\ 315\degr.  \citet{SLC01a} identified these bright new
features as \hsxii, \hsxiii\ and \hsxiv, while the flux at P.A.\
315\degr, although very suggestive,
is still too faint to reliably distinguish it from noise.

Comparison of the roughly concurrent STIS and ground-based \ion{He}{1}
images from early 2001 reveals at least some subset of the
population of young spots distributed about the ER which we inferred
from the \ion{He}{1} data.  Interpretation of the more recent
ground-based imaging from 2001 November is
ambiguous, given the high-resolution STIS antecedent.  It is possible, and
not without precedent, that additional hot spots have appeared,
resulting in the nearly-complete brightening about the ER seen in
Figure \ref{HeI}h.  The spots reported above have certainly brightened
as well, and at the ground-based resolution, spots separated by
20--30\degr\ will appear unresolved.  A spot's light curve can
accelerate quite dramatically in short periods of time, as we show in
\S\ref{ssec-evol}, thus it is possible that the extended flux seen
around \hsi\ in Figure \ref{HeI}h is due to sudden brightening of spots
10--12.  Questions such as these can only be resolved from {\em
HST} imaging, and given the rapid variation manifest in the number and
fluxes of hot spots, systematically-planned observations at {\em
least} twice a year are mandated, if we wish to properly monitor the
sequence of spot appearance and their evolution.  

%%%%%%%%%%%%%%%%%%%%%%%%%%%%%%%%%%%%%%%%%%%%%%%%%%%%%%%%%%%%%%%%%%

\subsection{Hot Spot Locations\label{ssec-loc}}

Using both the WFPC2 and STIS imaging data affords us a large set in
which we measure the positions of all currently detected hot spots.
We measure the flux center of the SN in the WFPC2 data using
the F675W image from 1994.  The wide passband ensures that
higher-velocity emission in the bright optical lines of H$\alpha$ and
[\ion{N}{2}] will be included, while at this early date the optical
ejecta was resolved but still relatively compact and appeared
symmetric.  We then geometrically registered this
image to the STIS data, and used the centroid in this frame for STIS
measurements.
Centroids for individual hot spots were measured using the PSF-fitting
routines within {\em daophot} applied to each difference
image in which the spot is unambiguously detected.
Using the positional uncertainties presented in \S\ref{sec-tests}, we
calculate the weighted average and weighted variance of the measured
centroids of each spot in each difference-image pair.
The resulting positions, relative to the measured SN centroid,
are presented in Table
\ref{tbl-pos}.  Quoted errors are the probable uncertainties in each
parameter, given by the weighted variance of each set of spot
positions, but do not take into account any systematic uncertainty
in our adopted SN centroid.  We will address this
in a forthcoming paper on astrometry of the SN
and surrounding field stars.  Preliminary results indicate that our
measured centroid is consistent with accurate astrometry of the SN in
1987 \citep{Rey95} using VLBI and {\em Hipparcos} positional data.

\begin{figure}
\centering
\epsscale{0.75}
%\plotone{../er.ps}
\plotone{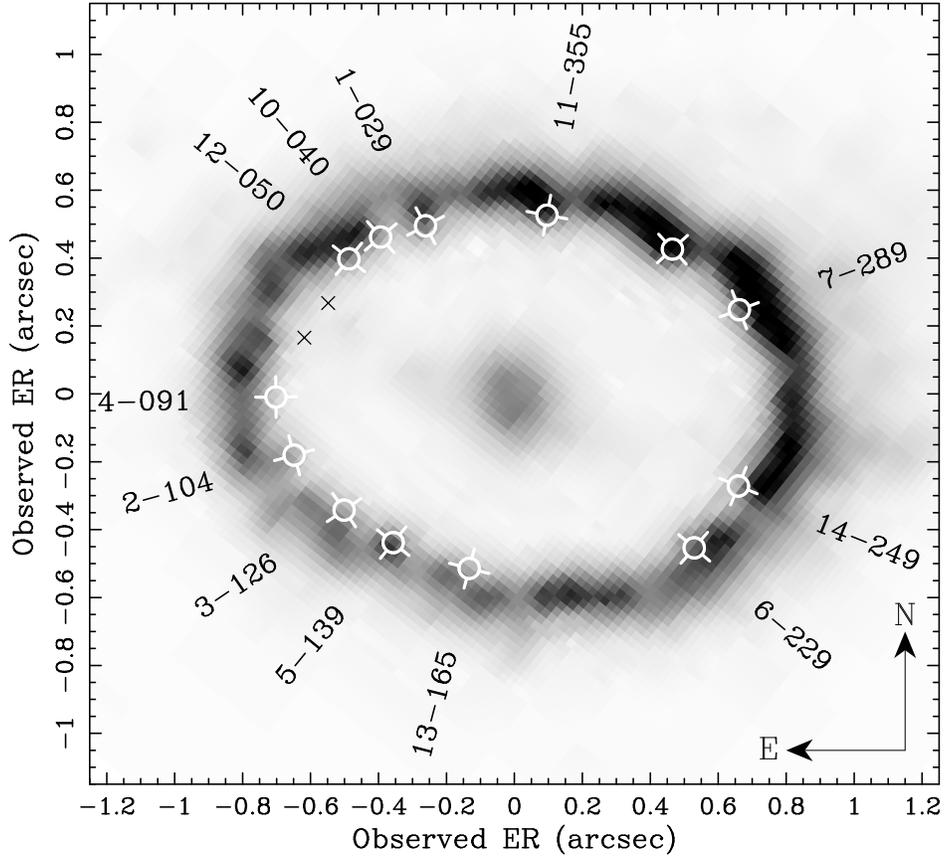}
\caption{WFPC2 F656N image
drizzled to double the PC chip resolution (\S\ref{ssec-param}) with the
positions and IDs of hot spots marked.  North is up and east is left.
Locations of brightening flux attributed to reverse-shocked ejecta are
marked with ``$\times$''.  The unlabeled cross-hair near P.A. 314\degr\ is
a marginally-detected spot (see text).
\label{er}}
\end{figure}

\placetable{tbl-pos}
\placefigure{er}

The orientation of the ER and locations of hot spots are shown
in Figure \ref{er}.  Spots are marked by cross-hairs, and confirmed
spots are labeled by their IDs.
As noted by \citet{SLC00}, the continued lack of brightening, and inward
radial positions of spots
\hsviii\ and \hsix\ appear more consistent with reverse-shock emission
than as ejecta-ER interaction (Paper I).  We mark their locations
by ``$\times$'' for clarity, but hereafter remove them from
the hot spot list.
Spot positions are also indicated around the ER in Figures
\ref{deproj}a and \ref{deproj}c.
As suggested by both {\em HST} and ground-based
\ion{He}{1} imaging, this system has undergone a
notable transition within the last year, from a few distinct hot spots at
isolated locations to a nearly continuous distribution around the ER,
with few breaks larger than 45\degr\ in P.A.

\begin{figure}
\centering
\epsscale{1.}
%\plotone{../hsmove.ps}
\plotone{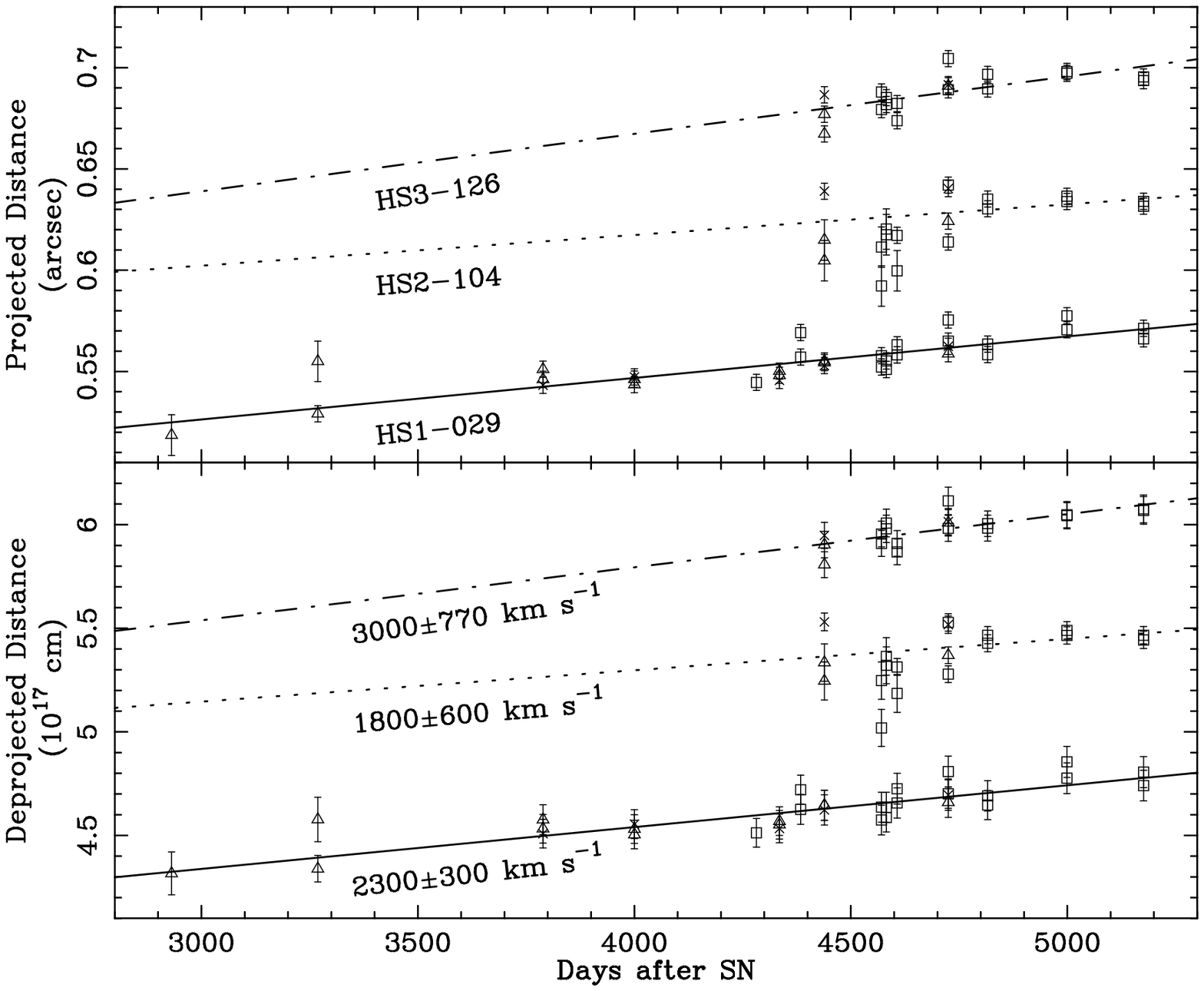}
\caption{Measured distance (top) and deprojected distance (bottom)
 from the SN of
 \hsi, \hsii, and \hsiii\ plotted versus time.  Deprojection was
 performed using the geometric parameters in Table \ref{tbl-geo}.
 For clarity, the positions of some spots have been shifted vertically: 
 \hsi\ by $-0.6\times10^{17}{\rm \ cm}$ (bottom), 
 \hsii\ by $-0\farcs04$ (top),
 \hsiii\ by $0\farcs08$ (top) and $0.5\times10^{17}{\rm \ cm}$ (bottom).
 Centroids were measured using the  PSF-fitting algorithms in {\em
 daophot}.  Error bars are $1\sigma$.
 STIS F28X50LP data are marked with  $\Box$, WFPC2 F656N data are marked
 with $\times$, and WFPC2 F675W  data are marked with $\triangle$.
 For each spot, the best-fit line (measured via linear least-squares
 fitting) is drawn through the data: solid lines for \hsi, dotted for
 \hsii, and dot-dashed for \hsiii.  The proper motion of the
 centroid of each hot spot is given by the slope of this line; the
 resulting velocities in the deprojected frame are listed along each
 line in the bottom panel, and in Table \ref{tbl-mov}.
\label{hsmove}}
\end{figure}

An obvious exercise to perform with this positional data is to search
for motion of hot spot centroids in time.  At the time of
writing this paper, only the first three hot spots have been
sufficiently bright for enough of an extended period of time to test
for proper motion.
We plot in Figure \ref{hsmove} the radial distance of these spots from
the SN as measured in filters containing H$\alpha$, for
those epochs in which each spot has an unambiguous profile for
centering.  The top half of each panel shows the observed (or
``projected,'' assuming the ER is a circular ring inclined to the line
of sight) radial position of each spot versus observation date.  The
bottom panel shows the deprojected distance of the spot from the SN,
using the geometric parameters determined in \S\ref{ssec-param}.  For
each locus of points, we show the best-fit line through the data,
determined by standard linear least-squares.  The slopes of these
lines give the proper motion of the centroid of each hot spot, and are
listed in Table \ref{tbl-mov}.  We denote the velocity of the centroid
by $v_{spot}$, which for the three spots listed have an average value
of 2000--3000 \kms.

\placefigure{hsmove}
\placetable{tbl-mov}

\begin{figure}
\centering
\epsscale{0.5}
%\plotone{../hsmove_err.ps}
\plotone{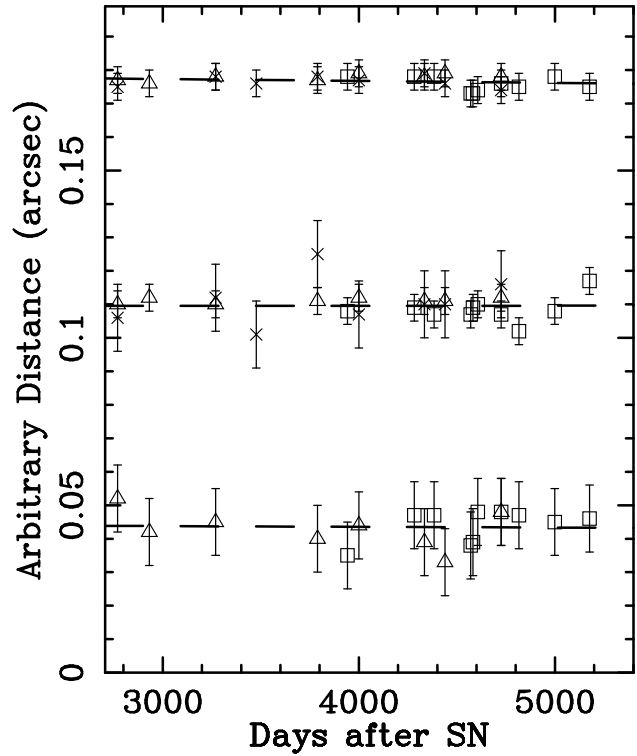}
\caption{Measured distance of three field stars versus time.  Symbols
are the same as Figure \ref{hsmove}.  A scalar has been removed from
each star's distance to position them such that stellar flux increases
with distance.   As expected, the positional scatter
increases as the flux of the star decreases.  Since F656N has a
narrower band-pass than the other two filters, the faintest star does
not show up in this filter.  The best-fit line to each star's data is
plotted, and all are consistent with zero proper-motion.
\label{hsmove_err}}
\end{figure}

Field stars within the LMC should not have significant proper motion
over the time period considered, thus the measured positions of these
stars serve as estimates of our expected random uncertainty.
We plot the radial distance of three stars (of different flux)
from the SN in Figure \ref{hsmove_err}.
Values along the $y$-axis have been offset by scalars such that
the brightest star is at the top of the figure.
Since the band-pass of the F28X50LP filter is wider
than that of F675W, which in turn is wider than F656N, a bright star
in the former appears faint in the latter.  As such, the faintest of
the three stars was not detectable in F656N.  We list in Table
\ref{tbl-mov} the slope of the best-fit line through each star's position,
and find these are consistent with no proper motion, as expected.  The
flux of Field Star 1 is similar to that of \hsi\ in F656N, but is far
brighter than any spot in F675W or F28X50LP; Field Star 2 is
representative of \hsi\ in the wide-pass filters, and a faint spot in
F656N; and Field Star 3 is similar to a faint spot in the
wide-band filters.  Using the above guidelines, \hsi\ is ``bright'' and
and \hsii\ and \hsiii\ are ``faint'' in most of the epochs
considered.  We see that the probable errors in the best-fit slopes of
the projected data are fairly consistent with a random noise model;
the deprojected errors are larger in part due to uncertainties in the
geometric parameters of the ER.

\placefigure{hsmove_err}

%%%%%%%%%%%%%%%%%%%%%%%%%%%%%%%%%%%%%%%%%%%%%%%%%%%%%%%%%%%%%%%%%%

\subsection{Geometric Parameters of the ER \label{ssec-param}}

To accurately interpret the positions of the hot spots on the ER, we begin
by studying the ER geometry.  A more detailed description of the
procedures will be presented in a forthcoming paper.  Briefly, we
generated an image of the ER at $0\farcs02278$ pix$^{-1}$ resolution
(double the PC-chip) by drizzling \citep{FR98} data at 2:1
resampling from the
F656N and F658N filters (H$\alpha$+[\ion{N}{2}]) between 1994 and 1996.
\hsi\ emission was low, but non-negligible in 1996, and we removed
its contributions to the ER by subtracting Tiny Tim model PSFs scaled
to the \hsi\ fluxes as determined in \S\ref{ssec-evol}.

The SN ejecta and
outer ring contributions were masked out, then Levenburg-Marquardt
least-squares minimization \citep{Press} was used to find the best-fit
ellipse to isophotes of the ER flux.  Since the flux distribution
in the ER is not uniform, but is generally brighter to the
northwest quadrant \citep{SLC01b,Pla95}, it is important that the flux
be normalized in a manner which does not give excess geometric weight to
any part of the ring, while still allowing the ER to be distinguished
from ``the shelf'' to the northeast \citep{Pla95} and low-level
surrounding flux.  We achieved this by
splitting the ER into $N$ wedges of equal angular size,
and normalizing the pixels in each wedge by the second-largest
pixel value in that wedge. This choice of normalization factor
was made to avoid contamination by bad pixels or
low-probability Gaussian outliers.  Isophotes of the ER were
constructed by including only the upper fraction $f$ of the normalized
flux from each wedge, where each wedge flux was integrated
in order of descending pixel value.
We fit a general ellipse with the center $(x_0,y_0)$,
axes $(a,b)$ and rotation angle $\phi$ as free parameters to the ER,
using isophotes for
$0.95 \le f \le 0.5$ in steps of 0.01, then averaged the resulting
parameters together for a variety of values of $N$.
These choices of $f$ correspond to the elliptical annuli containing
80\% and 20\% (respectively) of the total ER flux within 1\farcs6.
Assuming the ER is a circular
ring, inclined to the line of sight by $i$, then $i=\cos^{-1}b/a$.
The P.A.\ of the major axis is the rotation angle $\phi$.
The resulting values are listed in Table \ref{tbl-geo}, where reported
errors are variances only.    We will test the presence of systematics
within our method in a (aforementioned) forthcoming paper;  an
estimate of \onehalf\ pixel, or $\pm0\farcs011$  is suggested as a
combined uncertainty in axis lengths and SN centroid.

\placetable{tbl-geo}

\begin{figure}
\centering
\epsscale{1}
%\plotone{../deproj.ps}
\plotone{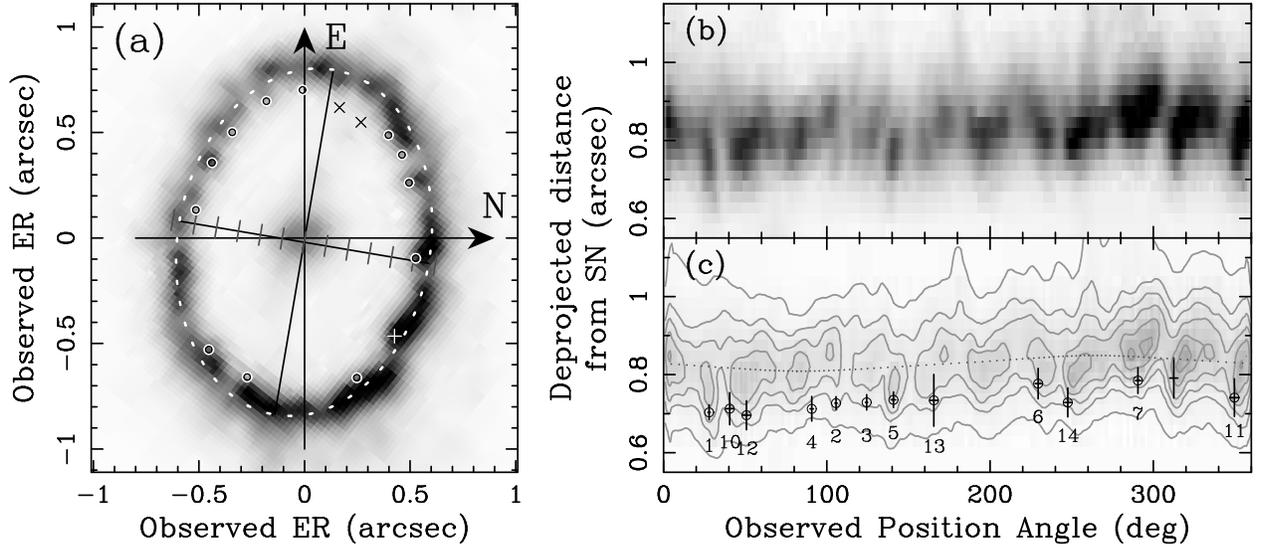}
\caption{(a) High resolution image of the ER from Figure \ref{er}
overlayed with the best-fit ellipsoid (dashed white line).
Major and minor axes are solid black
lines, and light travel delays of one month are marked with gray
major ticks along the minor axis.  The locations of hot spots are
marked by $\circ$ for confirmed spots and $+$ for the marginal detection.
(b) Panel (a)
has been radially deprojected using the best-fit ellipsoid axes and
orientation, but using the SN centroid, thereby plotting deprojected
distance from the SN against observed P.A.
(c) The same data as (b) with a lighter colormap and contours to
highlight structure.  The locations and positional uncertainties of
hot spots (Table \ref{tbl-pos}) as well as their IDs are indicated.
The dotted line indicates the position of the
best-fit ellipse, and appears as a sine wave due to the offset between
its centroid and the SN.
\label{deproj}}
\end{figure}

We compare our data with those from \citet{Pla95} and \citet{Bur95} in
Table \ref{tbl-geo}.   \citet{Pla95} performed a
careful study of the ER geometry using Pre-COSTAR WFPC and FOC images of
SN~1987A, predominantly in [\ion{O}{3}], but also including [\ion{N}{2}] at
lower signal-to-noise. \citet{Bur95} performed a similar study using
WFPC2 images in F502N, F547M, and F656N.
Our results are generally consistent with previous values.  Of
note, we find that the ER centroid is not coincident with our measured
value for that of the SN, as also found by \citet{Pla95}.
%The new semi-major axis is roughly the mean of smaller than previously
%found, and the ER may appear more rotated on the plane of the sky.
The high-resolution H$\alpha$+[\ion{N}{2}] image is shown in
Figure \ref{deproj}a, with the ellipse characterized by our best-fit
parameters drawn as a white dashed line.  The major and minor axes
have been drawn in black, showing the rotated P.A.\ of the major axis as
well as the offset centroid from that of the SN.
% (see \S\ref{ssec-loc}.
Tick marks in gray along the minor
axis denote light-travel delays of one month.  In general the ellipse
is a good fit,  however a detailed examination reveals considerable
scatter about this mean fit.   This is more easily seen
when the ER has been deprojected and unrolled into a map of
observed P.A.\ versus
rest-frame distance from the SN centroid, shown in Figure
\ref{deproj}b.  For clarity, a fainter stretch of the same data has been
plotted in Figure \ref{deproj}c, with contours to highlight
structure.  The dotted curve is the deprojected best-fit ellipse, and
takes on the sinusoidal shape since the SN and ER centroids do not
coincide.  A pure circular ring would map to this dotted line, and
we see considerable scatter about the mean.

\placefigure{deproj}

While it is somewhat simplistic to expect the ER, which is
clearly seen to be composed of bright knots embedded within tenuous
and extended gas, to be a perfect circular ring, it is unclear from this
mapping what its true geometry may be.  If the ER is inherently
elliptical [as suggested by \citet{Gou94,Gou98}], we would expect the
P.A.s of extremal recession velocity would not coincide with the
measured minor axes, while the nodes would not lie on the major axes;
this might be measurable by a detailed,
high-resolution (both spatial and in wavelength) spectrum.
Rather, if the ER were warped, sections could appear more or less
distant from the centroid than the majority of the gas, however this
effect could only be distinguished through three-dimensional mapping.
Currently, the three-dimensional circumstellar structures mapped in
light echoes by
\citet{CKH95} have not revealed such a warp in the equatorial
plane, however a re-analysis of that data using our improved techniques
of PSF-matched difference-imaging \citep{Sug02} might have the
sensitivity to reveal such features, if present.

\subsection{Deprojected Positions of the Hot Spots \label{ssec-deproj}}

To properly study the hot spot positions, they have been deprojected
using the ER geometry from Table \ref{tbl-geo}, assuming the ER is
intrinsically circular and that the SN is at a distance of 50 kpc.
Results are listed in Table \ref{tbl-pos} and 
have been plotted with their associated statistical uncertainties in
Figure \ref{deproj}c.
Also listed in Table \ref{tbl-pos} are $t_{earliest}$, the earliest epoch
at which the spot is unambiguously detected in {\em HST} images; $\Delta
t_{light}$, the number of days of light-delay between the spot and the
SN centroids;  
and $v_{blast}$, the blast velocity required to travel from the radial
distance of $0\farcs6$ at day 1300 to a
spot given its position and earliest detection, and including
corrections for light-travel delays.  This is the rough position and
epoch at which radio emission first appeared \citep{Man01}.
The large uncertainties in $v_{blast}$ result from large
distance to the SN combined with the small angular size of the ER.
We caution that this parameter greatly oversimplifies the system: we
assume a spherically-symmetric blast-wave expansion at early times,
that hot spots are all located within the plane of the ER, we use a
somewhat uncertain radial distance from super-resolved radio data,
and we ignore the variable cooling time ($t_c$, see \S\ref{ssec-gevol})
necessary for newly-shocked gas to cool into the optically-emitting
regime.  Velocities increase by 
\begin{equation}
 \Delta v_{blast} = \frac{8.68\times10^7 d_{50}(r_d-0\farcs6) t_c }
 {(t_{earliest}+\Delta t_{light}-1300^{\rm d}-t_c)^2} {\rm\ km\ s}^{-1}
 \label{eqn-1}
\end{equation}
where $r_d$ is the spot's deprojected position in arcseconds, $d_{50}$
is the distance to the SN in units of 50 kpc, and times
are given in days.   Quoted uncertainties in $v_{blast}$ to not
attempt to account for this unknown parameter, however we list in the
final column of Table \ref{tbl-pos} $\Delta v_{blast}$, the velocity
by which $v_{blast}$ would increase if $t_c=1$ yr is included.  While
$v_{blast}$ is not intended for quantitatively predictive purposes, it
is instructive when considering the required asymmetry of the system.

%%%%%%%%%%%%%%%%%%%%%%%%%%%%%%%%%%%%%%%%%%%%%%%%%%%%%%%%%%%%%%%%%%

\subsection{Evolution of the Hot Spots \label{ssec-evol}}

\begin{figure}
\centering
\epsscale{1.}
%\plotone{../ploths1.ps}
 \plotone{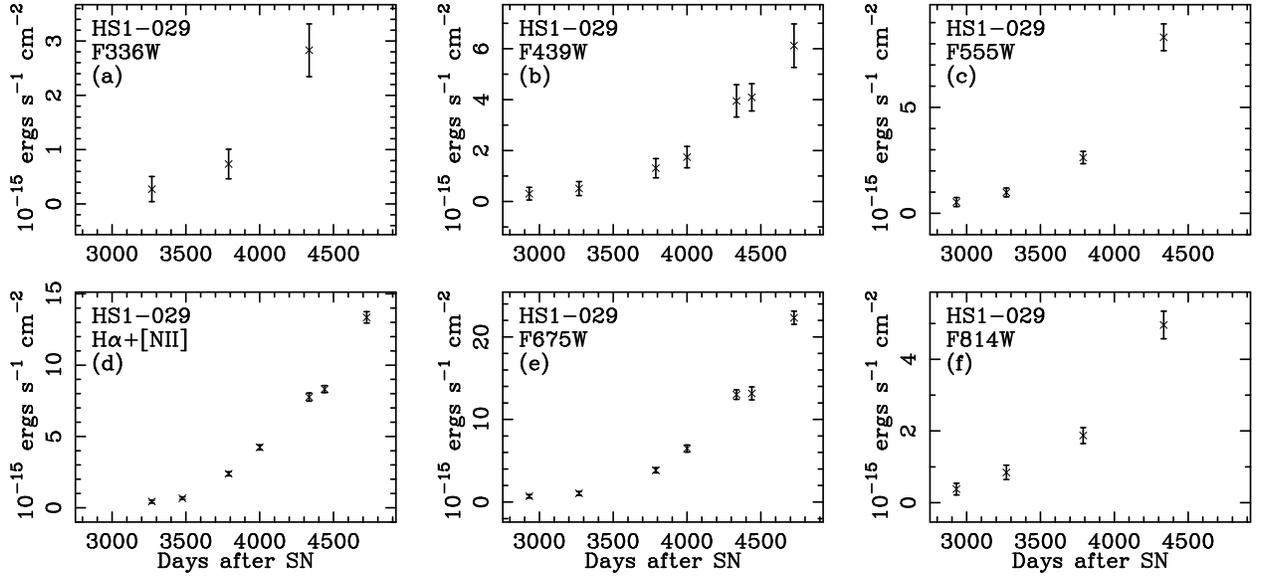}
 \caption{Light curves for \hsi\ in (a) F336W, (b) F439W, (c) F555W,
 (d) F656N \& F658N, (e) F675W, and (f) F814W.  Fluxes were measured
 using the PSF-fitting algorithms in {\em daophot}.  Error bars
 represent the formally propagated photon-count noise, sky uncertainty
 and fit errors.  Fluxes were calibrated using {\em synphot}, including
 decontamination corrections.
 \label{ploths1}}
\end{figure}

\begin{figure}
\centering
\epsscale{1}
% \plotone{../ploths1E.ps}
 \plotone{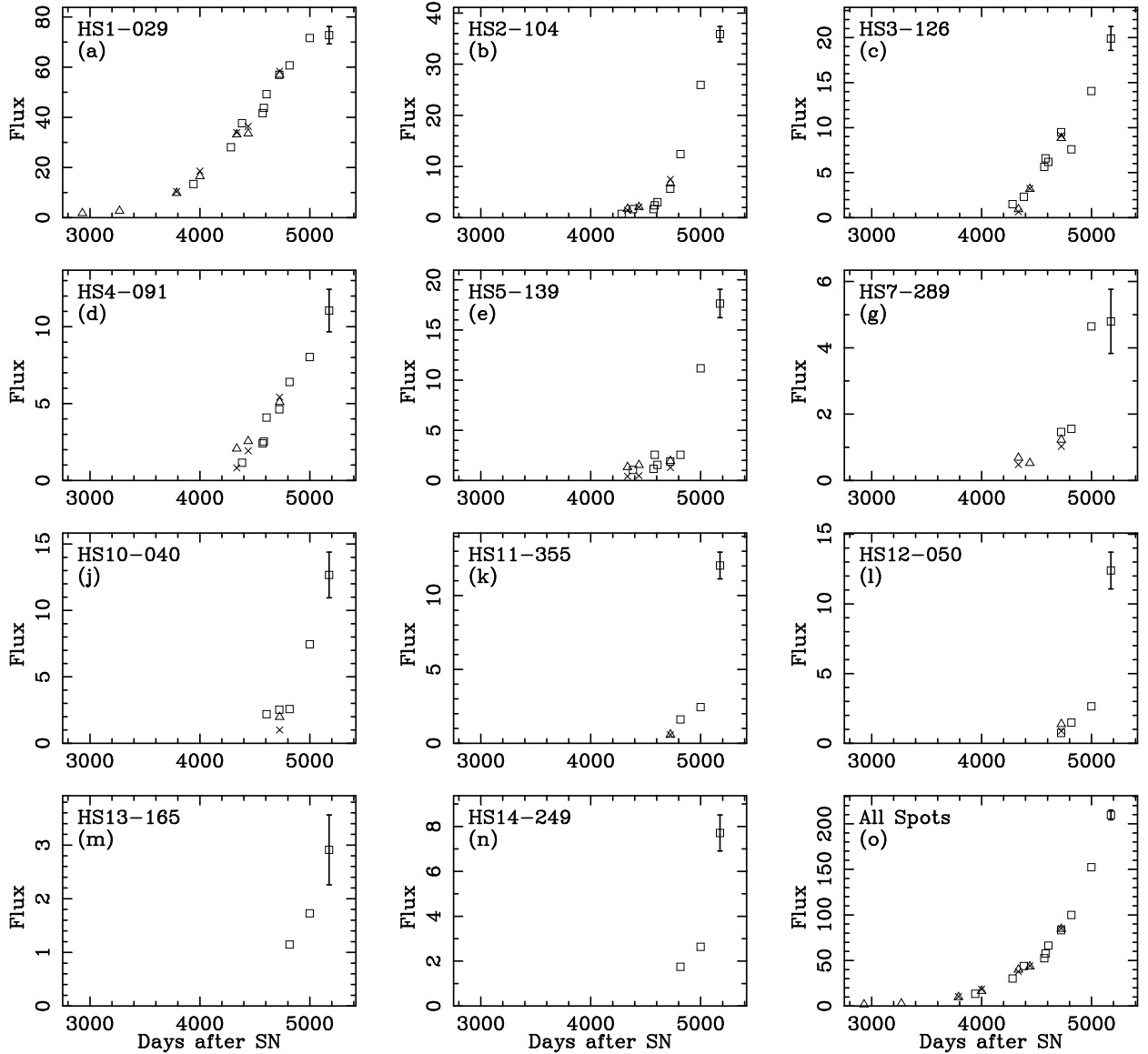}
 \caption{(a)--(n) Light curves for all confirmed hot spots except \hsvi\
 (which is confused with a coincidentally-aligned star) through {\em
 HST} filters containing H$\alpha$.
 Symbols are the same as Figure \ref{hsmove}.
 Calibrated WFPC2 fluxes have been
 empirically scaled to native STIS flux (counts sec$^{-1}$) using
 \hsi\ in 2000 February through both imagers:
 $F_{\rm F675W}=3.92\times 10^{-16}F_{\rm F28X50LP}$ and
 $F_{\rm F656N}=2.29\times 10^{-16}F_{\rm F28X50LP}$.  A
 representational $1-\sigma$ error bar has only been plotted on the
 last point in each panel for clarity.
 (o) The aggregate light curve of all known hot spot activity by summing the
 fluxes of each filter over all spots in panels (a)--(n).
 \label{ploths1E}}
\end{figure}

Although first reported in STIS spectroscopy from 1997 April, \hsi\ is
clearly detectable in 1996 in F502N, F555W, F675W, and F658N, and is
seen faintly in F555W and F675W in 1995 imaging
\citep[as noted by][and clearly demonstrated in Paper I]{Mic00a}.
A long time series of data exists in multiple WFPC2 wave bands for
\hsi, while a very well-sampled series of images are available for all
spots in the very broad STIS F28X50LP images.  We study the hot spot
evolution  by generating multi-wavelength light curves for all spots,
plotted in Figures \ref{ploths1} and \ref{ploths1E}.  All photometry was
performed using the PSF-fitting algorithm {\em allstar} in {\em daophot} (see
\S\ref{sec-tests}).  All error bars represent formally-propagated
photon-count noise, sky uncertainty and photometry error from
PSF-fitting, but ignore unknown systematics such as photometric calibrations.

\placefigure{ploths1}

We present in Figure \ref{ploths1} light curves for \hsi\ in WFPC2
filters.  Data points have been measured from images differenced from
the earliest epoch in each filter.  Panel (d)
displays the combined data from the F656N and F658N filters, both of
which contain the dominant H$\alpha$ hot spot emission.  The data
points from 1999 April (day 4440) in panels (b), (d) and (e) show an
apparent ``glitch'' in the light curve of \hsi, to which we return 
shortly.
Light curves in all filters have roughly similar evolution in spite of
the different emission lines contributing to each wavelength range.
Using an implementation of the Levenberg-Marquardt non-linear
least-squares minimization \citep{Press}, we fit simple analytic
functions to these curves in an attempt to quantify this first hot
spot's evolution.  A power law of the form
$A(t-t_0)^b$ describes the data well, however a wide range of
parameters offer equally satisfactory fits, since e.g., variation of $A$
compensates for
decreasing $t_0$ and increasing index $b$.  We find $b\sim5.5$ yields
reasonable fits for all filters, but requires $t_0\sim$500--2000 days.
If $t_0$
represents the initial turn on of the hot spot, these small values are
unphysical.  If we set $t_0=2500$, corresponding to the earliest WFPC2
observation, we find $b\sim3$--3.5.
A simple exponential $Ae^{bt}$ fits all the data well with
$b\sim\;$(1.3--1.7)$\times10^{-3} {\rm day}^{-1}$, as does a standard Gaussian
$Ae^{-(t-t_0)^2/\sigma^2}$, with parameters $t_0\sim$5400--5600 days
and $\sigma\sim$1000--1300 days.

\placefigure{ploths1E}

In Figure \ref{ploths1E}a--n, we plot the light curves through filters
containing H$\alpha$, for all confirmed spots except \hsvi.  An
accurate flux for this spot is difficult to measure since it is
roughly coincident with star 5; difference images scaled to
the ring flux leave residuals of Star 5, and photometrically-scaled
difference images do not compensate for the fading ring flux.
Since \hsi\ turned on before the earliest F28X50LP observation by STIS
in 1997 December, we measure the 1997 flux in this spot by differencing the
earliest STIS image from the 1994 September PC image in F675W,
after having been been geometrically registered to the STIS plate scale.
All \hsi\ fluxes from F28X50LP have had this flux from 1997 added back
in.  All other data points represent averages of the fluxes measured
from images differenced from multiple epochs before a given spot
turned on.  To overplot fluxes from three different filters, we
determined an empirical scaling between F28X50LP flux (in counts sec$^{-1}$)
and calibrated WFPC2 fluxes (in \ergcms) using
\hsi\ measured in 2000 February by both imagers.  For
clarity, a representational $1\sigma$ error bar is only plotted on the
last data point in each panel.

\begin{figure}
\centering
\epsscale{0.5}
% \plotone{../stable.ps}
 \plotone{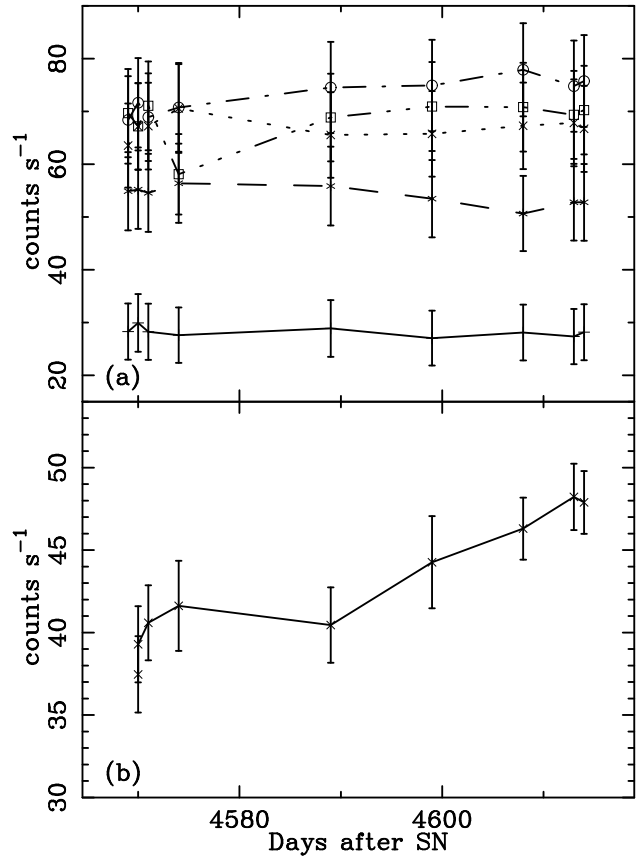}
 \caption{(a) Light curves for five stars of flux comparable to \hsi\
 between 1999 August -- October through STIS F28X50LP using individual
 cosmic-ray split pairs of images.  We see no
 evidence for calibration errors between individual epochs.
 (b) Light curve for \hsi\ in the same data.  The hot spot evolves on
 timescales of $\lesssim 1$ month.  The rapid variation in the first
 three points is likely due to bad pixel contamination.
 \label{stable}}
\end{figure}

Examination of panel (a) indicates that a short pause did occur after
1999 February, during which the hot spot flux remained depressed from
an extrapolated brightening profile until
early 1999 October.  To ensure that this is not
due to calibration errors, we found field stars with fluxes similar to that
of \hsi\ and which are constant to within 2.5\%, as measured in data
exclusive of 1999 August--October.  Photometry for five of these stars
for individual cosmic-ray split pairs of images between 1999
August--October is shown in
Figure \ref{stable}a.  If the glitch resulted from calibration errors in
the data, these light curves would reflect the same trend.  That the
curves are constant indicates that any variation in the data from this
period is real.  To test for short-term ``flickering'' in \hsi, we
plot in Figure \ref{stable}b the light curve for the hot spot,
generated from the same individual
observations as panel (a), differenced from the 1997 December STIS image.
We see little  
evidence for short-term variations aside from a marginally-significant
change during the 27 hours separating the first 3 data points, which
we trace to bad-pixel contamination of the hot spot in one image.

\placefigure{stable}

The light curve for \hsi\ appears to be changing its slope, and perhaps
reaching a plateau, in the most
recent epochs.  This curve is visually suggestive of a Gaussian
function, and is best fit with $t_0=$5430--5550 days,
$\sigma=$1100--1200 days, which should imply a
maximum around 2002 January--March, {\em if} the brightening profile is
Gaussian.  These parameters are quite consistent with
those derived from the WFPC2-only light curves, which did not contain
the last three data points showing the break in slope.   We re-address
this in \S\ref{ssec-gevol}.

Other hot spots appear to evolve along one of two rough
classifications.  \hsiii\ and \hsiv\ brighten almost linearly in
time from the earliest detection, while \hsii\ and \hsv\ accelerate
markedly after a period of slow
growth, describing a somewhat more exponential evolution.
%As shown in Figure \ref{deproj}c, spots such as \hsi,
%\hsii, \hsv, \hsvi\ and \hsxi\ lie on steeper flux gradients than
%e.g.\ spots \hsiii\ and \hsiv, suggesting that the former group of
%spots brighten more rapidly in time, if the flux gradient is assumed
%to be an indication of relative density.
If \hsi\ is an indication of general hot spot evolution, each light
curve could plateau after an as-yet-unknown period of growth.
The evolution of \hsvii\ is somewhat unclear since
its most recent datum is depressed relative to either linear or
exponential growth, suggesting a possible break similar
to, but much more rapid than, \hsi.  Although much younger than the
first four spots, \hsxi, \hsxii, and \hsxiv\ currently appear to
follow the exponential growth pattern while \hsx\ could be increasing
linearly; \hsxiii\ (and realistically, most of these newer spots) is
too faint for a reasonable assessment as of 2001 April.  
%Gaussian fits to most spots (other than \hsi) show
%them reaching peak brightness almost immediately after the last epoch
%of data, which we believe reflects a straightforward numerical
%difficulty with the function, rather than a physical prediction.
%A Gaussian has roughly constant slope near its inflection points,
%thus in the absence of a significant change in slope (e.g.\ as it
%approaches its peak), a generalized Gaussian with a free zero-point in
%the abscissa can fit moderately linear data with a variety of
%disparate parameter sets.
%We caution that while a Gaussian appears to fit these early data,
%there is no  {\em a priori} reason to expect the long-term evolution
%of a light-curve to follow this function, thus these  functional fits
%do not currently  serve any predictive capacity.

%%%%%%%%%%%%%%%%%%%%%%%%%%%%%%%%%%%%%%%%%%%%%%%%%%%%%%%%%%%%%%%%%%

\subsection{Aggregate Evolution of the ER-Ejecta Interaction \label{ssec-bulk}}

\begin{figure}
\centering
\epsscale{0.5}
% \plotone{../HeI_plot.ps}
 \plotone{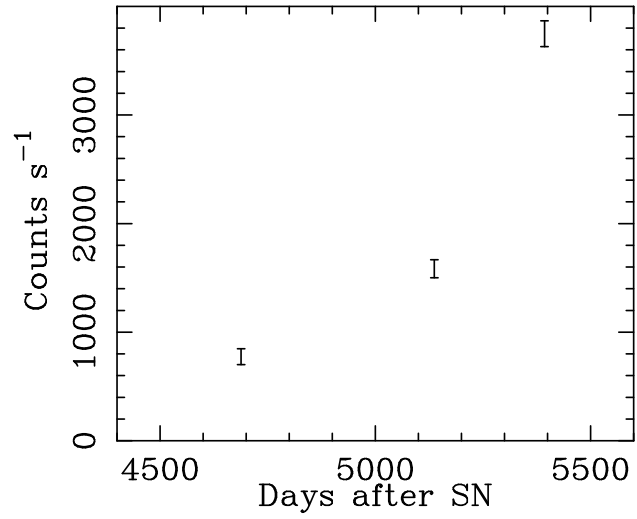}
\caption{Light curve of the increase in \ion{He}{1} emission in SNR
1987A since 1998; since \hsi\ was already a source of line emission at
this epoch, this light curve contains an undetermined (but small)
zero-point.  Data have not been photometrically calibrated, hence
we present it in units of instrumental counts per second.
\label{HeI_plot}}
\end{figure}

The sum of the light curves in Figure \ref{ploths1E}a--n
gives the aggregate flux of the ER-ejecta interaction.  We plot
this total spot flux in each filter in Figure \ref{ploths1E}o; this
would be the observed light curve (corrected for the fading underlying
ER material) if the SNR were fully unresolved in our imaging, i.e.\ as is
typical of ground-based imaging.  In Figure \ref{HeI_plot}, we
plot total spot flux in \ion{He}{1} (measured from data shown in
Figure \ref{HeI}).  This light curve has an undetermined zero-point,
since we have measured the change in \ion{He}{1} flux since 1998, at
which time \hsi\ was already a source of line emission.  Since these
data have not yet been photometrically calibrated, we present them in
units of instrumental counts per second.

\placefigure{HeI_plot}

\citet{Suz93}, \citet{Mas94}, and \citet{BBM97b} have modeled X-ray
light curves for the
ER-ejecta interaction, but did not address soft UV or optical
transitions.  \citet{Luo91} modeled the impact in the soft UV, but
only \citet{Luo94} directly addressed the expected optical emission,
by assuming  isotropic ejecta impact an idealized toroidal ring.
Using the empirical scaling between F28X50LP and F656N, the total
flux in all spots has grown monotonically to $5\times 10^{-14}$
\ergcms\ in roughly
six years.  Since the hot spot emission is dominated by H$\alpha$, we
compare this value to the \citet{Luo94} models, which reach the
same  line flux far more rapidly:  0.25, 0.75, 0.7, and 0.65 years
after the line emission begins (for density models A, B, C, and D
respectively).
%These model light curves all turn on immediately, whereas
%the actual evolution increases quite gradually.
Following a nearly instantaneous turn-on, models A, B, and D grow
roughly linearly to a peak and turn over in approximately 1.5, 3.5 and
3 years, respectively, while model C is still brightening (in bursts)
after 6.5 years.  In contrast, the observed evolution at the time of
publication is
quite gradual and has yet to plateau.  Finally,
the predicted model flux for the latest data reported in this paper
should be between $10^{-13}$--$10^{-12}$ \ergcms, having already
passed the maximum peak flux. It is not surprising that this model
poorly correlates to the observed evolution, since the actual
interaction has proven stochastic while their idealized model
considered the impact simultaneous along the whole inner surface of a
toroidal ER.   Furthermore, their model did not consider the blast
front propagating through an intervening \ion{H}{2} region.
It remains to be seen whether the integrated light curve of the
ER-ejecta interaction will evolve like one of these models once the
ejecta impact the main inward surface of the ring.

%%%%%%%%%%%%%%%%%%%%%%%%%%%%%%%%%%%%%%%%%%%%%%%%%%%%%%%%%%%%%%%%%%

\section{Consistency and Completeness  \label{sec-tests}}

As demonstrated by \citet{Law00} and noted by \citet{MSP00},
hot spots are most easily detected at early times in spectra,
however
in the absence of a systematically-planned STIS observing program
which samples the entire ER (such as that presented in this work),
they are  most effectively revealed with PSF-matched difference
imaging. As noted in \S\ref{sec-obs}, hot pixels constitute a
significant source of contamination and confusion in {\em HST} data,
and the accurate
detection of new hot spots at early times depends entirely on our ability
to distinguish them from the abundant sources of noise.
Proper detection of this rapidly-changing phenomenon requires
confirmation in multiple epochs of data,
however high-resolution data are relatively sparse: WFPC2 imaging is
taken roughly only once per year, and STIS spectra are predominantly
narrow-slit slices through the ER rather than 2\arcsec\ spectral
images of the entire ring.
%; fortunately, STIS data from 2000 and 2001
%was not subject to the standard 12-month proprietary period.
While we make every effort to maximize the data quality of
{\em HST} imaging to reliably detect new loci of interaction along the ER,
it is important to understand the limitations inherent to these data.

To test the reliability of new hot spot detection, we generated
artificial data as follows.
All known hot spots were removed from the F656N images from 1999
January and 1999 April using a {\em Tiny Tim} model PSF, then 10 pairs of
images were created, each
containing a random number of hot spots (average of 5 spots per
image), of random flux (ranging from the faintest to brightest actual
spot fluxes), situated 
randomly around the inner-half of the ER (where we observe new spots
forming), with the flux of each spot in 1999 April 10\% brighter than
1999 January.  These data were differenced from F656N images from 1998
and 1997, and analyzed using the same criteria as applied to actual
spot detection.   Table \ref{tbl-acc} contains the results from this
exercise.  For each given flux range, $N_{\rm Detected}$ is the total number
of sources we identified in all images, $N_{\rm False}$ is the number
of sources which were not actual spots, and $N_{\rm Missed}$ is the
number of actual spots which we missed.  We see that our detection
procedure is
$\sim$100\% complete for the majority of the applicable flux range, and
roughly 60\% complete at the faintest end, corresponding to the very
faintest spots (e.g.\ \hsv) at their earliest detections.

\placetable{tbl-acc}

The employment of PSF-fitting crowded-field photometric
techniques from {\em daophot} follows from the obvious crowding in
some hot spot regions,
but also from the noise characteristics of the data. As seen in
Figure \ref{images}, hot spots lie on the inner edge of the ER,
translating in difference images to the boundary between the
uniform region interior to the ring, and the non-uniformly fading ER.
Since the background varies on pixel-to-pixel scales, a large sky annulus
will likely not reflect the actual background of the
hot spot.  Many hot spots are faint, and are more strongly affected by
Poisson noise and the omnipresent warm pixel contamination when
performing simple aperture photometry.  The {\em allstar} task
performs a non-linear least-squares fit of an analytic stellar profile to
each source, while dynamically fitting both the underlying sky and
nearby sources.  This should, in principle, provide the most reliable
centroid and flux estimate for a hot spot, by directly addressing the
above difficulties.

This was directly tested by generating an additional set of 5 pairs of
images with hot spots of random flux placed at the known hot spot
positions, differencing them as above, and performing both standard
aperture (with {\em phot}), and crowded-field (with {\em allstar})
photometry.  Photometric accuracy for the latter, and positional
accuracy for both algorithms are listed in the final three columns of
Table \ref{tbl-acc}.  Aperture photometry yielded
errors of roughly 50\% for fainter sources and in crowded regions.  In
contrast, {\em allstar} errors are generally within the formal
photometric uncertainties.  In most cases,
centroids measured with {\em allstar} are more accurate as well, since
flux-weighted centroids often drift toward a brighter nearby source or warm
pixel.

%%%%%%%%%%%%%%%%%%%%%%%%%%%%%%%%%%%%%%%%%%%%%%%%%%%%%%%%%%%%%%%%%%

\section{DISCUSSION \label{sec-disc}}

\subsection{Generalized Hot Spot Evolution \label{ssec-gevol}}

Following the radiative shock model of \citet[][and references
therein]{Mic00}, the SN ejecta expanded at $\lesssim3\times10^4$ \kms\
until it encountered an \ion{H}{2} region of density $n_{\rm
H\,\mbox{\tiny II}}$ interior to the ER, which was formed when 
the blue supergiant progenitor photoionized  stellar
winds ejected during its previous red supergiant phase \citep{CD95}.
This encounter drove a forward shock into the \ion{H}{2} region with a
velocity $v_b$, which ultimately impacts a protrusion of density $n$
in the ER.  The density jump implies the shock propagates into the
protrusion with velocity 
$v_s=v_b f(\theta)\sqrt{n_{\rm H\,\mbox{\tiny II}}/n}$ 
where $f(\theta)$ accounts for the obliquity of the shock\footnote{
These values of $f(\theta)$ were calculated for a plane-parallel shock
impacting a cylinder with density contrast
$n/n_{\rm H\,\mbox{\tiny II}} = 70$, and ignored time-variable
effects (E.~Michael, private communication).},
and ranges from 2 for a head-on shock to 0.7 for
$\theta=\frac{\pi}{2}$. 
  The time for the shocked
gas to radiatively cool from the postshock $\sim10^6$~K to
$\sim10^4$~K, at which the gas emits largely in the optical, is
$t_c$. For velocities in the 
range of a few hundred \kms, \citet{Mic00} report
$t_c=4.6n_4^{-1}\left({v_s}/{300 
{\rm\ km\ s}^{-1}} \right)^{3.7}$ yr, where $n_4$ is the density in
units of $10^4$ cm$^{-3}$.
With infinite spatial resolution, the optical observation of a new
parcel of emitting gas would only show where the shock had been a time
$t_c$ earlier.
Since this cooling time is a function of $v_s$, $n$ and the chemical
composition of the ER, each emitting parcel of gas does so with its
own set of these parameters.
The hot spots which we observe are the integrated light
curves of all the ER material in a seeing element through which the
forward shock has passed, and the centroids we measure,
derived from regions too small to be resolved by {\em HST}, are the
luminosity-weighted 2-D projections of 3-D emitting volumes.
The observed motion of the hot spot's centroids is thus the
measurement of the radial increase in {\em average} emitting area in
time, and can only serve as a lower limit to the velocity of the shock 
causing the emission.

Consider what one would observe as the forward blast moves
down the side of an inward-facing protrusion.  If one idealizes the
protrusion as a cylinder of constant density, the forward shock
as planar with constant velocity $v_b$, and ignoring optical thickness
effects and radiative precursors, then the cooling time $t_c$ will be
constant. As the shock moves down the cylinder, the transmitted shock
velocity $v_s$ will
be constant and much less than $v_b$, thus an observation at infinite
resolution would show a uniform growth of emission down the axis at
velocity $v_b$.  At our limited resolution, we instead observe the
flux-barycenter of all emitting material, and hence measure a
velocity equal to $\case{1}{2}v_b$.  For a similarly constructed
conical (or truncated cone) protrusion, the observed velocity would be
$\case{2}{3}v_b$.
Of course, an actual protrusion has a complicated
(and as yet unknown) geometry and density structure, both of which
mediate the variation of the cooling timescale and the mass (per unit
time) swept-up by the forward blast, as the blast moves down the
protrusion.  Nonetheless, from geometry alone we expect the observed
spot velocity to be of order half the blast velocity.
From Table \ref{tbl-pos}, $v_{blast}$ is the inferred blast velocity
needed to reach a given spot, assuming the blast expanded uniformly
to $0\farcs6$ during the first $\sim 1300$ days, and has an average
value of $\sim$3000--5000 \kms for most spots.  
Recall that these values do not include estimates for $t_c$, which
would increase these inferred velocities by hundreds of \kms\ per year
of cooling time (Eq.\ \ref{eqn-1}).
Assuming a 1--2 yr cooling time, we find the values of $v_{blast}$
are of order 1--2 times those of $v_{spot}$, consistent with the
heuristic expectation above.  We thus interpret $v_{spot}$ as a rough
measurement of half the blast velocity, implying $v_b\sim3000-5000$
\kms. 

Hot spot evolution is now considered according to the following
scenario.  A hot spot arises on an inward-facing protrusion from the
ER, as shown in Figure 3 of \citet{Mic00}.  From the radiative shock
model, $t_c \propto f(\theta)^{3.7}$ thus the time at which
post-shocked gas actually emits visible line radiation is a sensitive
function of the obliquity of the shock.
The blast wave first strikes the tip of the protrusion head-on, and
thus has a long cooling time (since $f(0)\sim~2$).  As the blast
wave travels down the length of the protrusion, $f(\theta)\rightarrow
0.7$ hence $t_c$ decreases substantially.  For $v_b=3000$ \kms\ and
$n=10^4 {\rm\ cm}^{-3}$, $t_c\sim60$ yr for a head-on shock, and
$t_c\sim1$ yr for a side shock.  An observed hot spot at
early times is thus the shocked gas along the protrusion's sides
cooling into an optically-emitting temperature range.  Once the shock
reaches the base of the protrusion, $f(\theta)\rightarrow2$ and $t_c$
increases again.  Flux in the early-time light curve will grow
substantially as new gas is shocked by the blast wave along the side
of the protrusion.
Once the blast reaches the protrusion base, emission will be dominated
by newly-shocked gas from the slow shocks moving into the protrusion along
its sides, and since roughly equal volume will be swept-up per unit
time, the light curve should flatten and grow monotonically (assuming
the shocked gas does not cool out of the optical for a long time,
perhaps due to photoionization from shocked material upstream).
Only after many years will the light curve suddenly accelerate again, as
the gas at the tip and base cool into the optical-emitting regime.
At early times, the tips of the shocked protrusions should be
substantial X-ray emitters, while the sides emit optical and near-IR
lines.  The correlation of a population of X-ray bright spots with
optical hot spots \citep{Par02} in {\em Chandra} imaging is suggestive
of this interpretation.

The late-time flattening of the lightcurve of \hsi\
may be explained via this model.  From Figure \ref{deproj},
the protrusion on which \hsi\ appears is long and fairly isolated
from the neighboring ER material.  Idealizing this as a truncated
conical protrusion of length $0\farcs05$--$0\farcs1$ (as measured in
panel b of Figure \ref{deproj}), the forward blast would require
roughly 1500--3000 days 
to travel its length, assuming a velocity of 3000 \kms
(\S\ref{ssec-loc}).  As shown in \S\ref{ssec-evol}, the light-curve of
\hsi\ shows a significant change in slope beginning sometime around
day 4700--5000, i.e.\ 1800--2200 days after its earliest appearance,
and within the expected range noted above.  
A physical scale of \hsi\ can be estimated using the radial position of
\hsx, assuming the forward shock had to pass the \hsi\ material before
impacting that of \hsx.  The radial position of \hsx\ is highly
uncertain; however, using its central value and the initial radial
distance of \hsi, the deprojected radial difference in
distance is $\sim0\farcs05$, or $3.8d_{50}\times 10^{16}$ cm.  This is
consistent with the value of $3\times10^{16}$ cm found by
\citet{Mic00} using spectroscopic considerations, and at this size, the
expected time for the break in slope of the light curve of \hsi\ is
consistent with that observed.  This is suggestive that the heuristics
of our model are correct, and that the forward blast is nearing the inner
edge of the ER (rather than a protrusion) in the vicinity of \hsi.

Our model also suggests a possible explanation for the linear light
curves of spots 2--5.  If these protrusions are small in
radial extent, the time necessary for the shock to move down their sides is
greatly reduced, hence the spot's emission would spend little time in
the  initial rapid-brightening phase, and would quickly display the
more monotonic increase in flux consistent with the transmitted shock
moving into its sides.   If this interpretation is correct, the
observed light-curves of the first five spots thus suggest that \hsi\
lies on a longer radial protrusion than that of \hsii\ or \hsv,
which in turn are longer than \hsiii\ or \hsiv.

%into the base of the emitting protrusion.   First appearing around day
%2900, and using $v_{blast}=2900$ \kms, we find a size of the \hsi\
%protrusion of $4.5--5.3 d_{50}\times 10^16$ cm, where where $d_{50}$
%is the distance to the SN in units of 50 kpc.  We can

%%%%%%%%%%%%%%%%%%%%%%%%%%%%%%%%%%%%%%%%%%%%%%%%%%%%%%%%%%%%%%%%%%

\subsection{Hot Spot Timing\label{ssec-time}}

Why did \hsi\ turn on over three years prior to any others?  Is there
something unique about this spot which requires a separate physical
explanation?
Figure \ref{er} suggests that hot spots occur on inward-facing
protrusions from the main ER, which would naturally be the first sites
to be struck by an isotropically-expanding forward blast wave.
From Figure \ref{deproj}, we see that nearly all spots do appear to
lie at the extrema of inward-facing protrusions.
Furthermore, \hsi\ is located, in
the deprojected frame, on the most inward-facing protrusion,
and from Figure \ref{hsmove} we see that its early-time position was
roughly $0\farcs06$ closer to the SN than other spots.  Examination of
the values of $v_{blast}$ in Table \ref{tbl-pos}, which were
calculated using the early-time positions of spots, suggests that
there was no preferential velocity required for \hsi\ to turn on as
early as it did.  Rather, we argue that the early appearance of \hsi\
results mainly from its inward position with respect to the other hot
spot protrusions.

Values of $v_{blast}$ for all spots are suggestive of a roughly
uniform blast velocity
of 3000--5000 \kms, consistent with that inferred in the radio by
\citet{Man01}, in {\em HST} spectroscopy by \citet{Mic00}, in {\em
Chandra} X-ray imaging ($v\sim5200\pm2100$ \kms) by \citet{Par02}
and in {\em Chandra} X-ray spectra ($v\sim3400\pm700$ \kms) by
\citet{Mic01}.
We now explore whether we can glean additional information about the
forward blast from the observed positions and earliest appearance of
the hot spots.
If we use $t_{earliest}+\Delta t_{light}$ as a relative
estimate (with respect to the SN) of when a hot spot first turned on,
the following spots are mutually coeval:
spots 2 and 3; spots 4 through 6; spots 10 and 13; and spots 11, 12,
and 14. Since  \hsii\ and \hsiii\ have nearly identical deprojected
distances and values of $v_{blast}$, there is no need to invoke
significant inhomogeneity in the forward blast to explain the order of
the appearance of these neighboring spots.
Spots 4--6, however, suggest a different trend.  The deprojected
distances of \hsiv\ and \hsvi\ are marginally discrepant at the
1$\sigma$ level, implying that the blast wave had to travel
significantly faster to \hsvi\ for it to turn on at the same time as
\hsiv.  Although the uncertainties are large, hot spot positions
suggest a rough trend that the required shock velocity 
varies somewhat continuously with P.A., lowest to the north and
northeast around \hsxi\ and \hsi, and increasing to the west around
\hsvi\ and \hsvii.  This could also be explained by invoking
variations in shapes of the protrusions hosting the hot spots, since
the cooling time $t_c$ is such a strong function of geometry.  

Inspection of the locations of spots around the ER shows that roughly
$\frac{3}{4}$ of confirmed spots are located along the eastern half of
the ring.  This would be the natural result were the forward blast to
arrive at that half first.  This suggests two possible explanations.
First, the SN centroid is offset to the east of that of the ER.
Recall that in \S\ref{ssec-param}, we find our measured SN centroid is
offset by 20 mas to the east of the ER centroid.  We also find in
Table \ref{tbl-pos} that eastern spots appear closer (in the
deprojected frame) to the SN than those in the west.  
A simple assumption of symmetry suggests that the deprojected
positions of hot spots should be roughly equidistant from the SN,
implying that our measured SN centroid might be inaccurate, perhaps
due to asymmetry in the ejecta in 1994, or patchy absorption from dust
within the ejecta (as suggested by our referee).  
We quickly tested this by measuring spot positions from the ER 
centroid, and found that while
this does result in an increase of the deprojected distance of
easterly spots and a decrease for westerly spots, the overall
effect is not large enough to fully reconcile the discrepant distances
presented.  Recalling that spots appear on inward-facing protrusions,
we may speculate that in addition to an offset SN centroid, spots to
the east may lie on systematically-longer protrusions.  With the data
currently presented, we can not rule out this hypothesis or the first
explanation.  

Second, an inhomogeneity in the forward-blast exists on
the largest scale. 
\citet{Mic00a} find from narrow-slit STIS spectra that the reverse
shock is roughly 5\% further from the SN at P.A.$\sim$220\degr\ than at
P.A.$\sim$40\degr, concluding that CSM interior to the ER is
more tenuous along the far (southwestern) side of the ring.
Recent images 	of the radio \citep{Man01} and X-ray \citep{Bur00}
remnants both show increased emission toward the eastern
side of the ER.  \citet{Bur00} interprets the bulk of the X-ray and
non-thermal radio emission as arising from a zone of shocked SN debris
and CSM between the forward blast and reverse shock.  A
detailed study of {\em Chandra} X-ray imaging \citep{Par02} shows that
X-ray bright knots are well correlated with the optical hot spots,
however \citet{Mic01} argue that only $\sim$4\% of the X-ray emission
can come from these spots, and suggest that the observed X-ray
asymmetry is  indicative of denser CSM material to the east interacting
with the blast wave.  This, they propose, could result from either an
asymmetric SN explosion or an asymmetric CSM, as follows.  If the SN
explosion is symmetric, and the CSM is denser to the east, then the
forward blast would travel faster into the western half of the ER.
Conversely, if the CSM is symmetric, then the blast wave must have
travelled faster in the eastern direction, shocking denser CSM closer
to the ER. 

We briefly consider this latter scenario.  If the increased radio
emission to the east results uniquely from a higher blast velocity
toward that half of the ring, we should expect the western emission to
mimic that in the east with a roughly constant time delay.  
Comparison of super-resolved images of SNR~1987A from \citet{Man01}
show that the western radio emission in 1999.7 and 2000.8 is at
roughly the same flux as eastern emission in 1995.7 and 1996.7,
respectively.  Continued monitoring of the radio remnant will show
whether this four-year time-delay is persistent.  

Consider now the former scenario, in which the CSM is asymmetric.
The three-ring nebulosity surrounding the SN has largely been
attributed to the interaction of the progenitor star's blue and red
supergiant (BSG and RSG) winds via the interacting stellar winds model
\citep{Kwo82,Bal87}.
Equatorial overdensities in a previously-expelled slow, dense wind
focus a fast, tenuous wind into a polar trajectory, the interaction
from which produces a bipolar peanut-shaped nebula, or wind-blown bubble
\citep{Woo88,Arn89,CE89,Luo91,WM92}.   The ER is the overdense waist of
the peanut, the inside of which was largely evacuated by the BSG
wind.  The visible ring is most likely the inner skin of the
overdensity, which was optically thick to the UV and soft X-ray flash
from the SN, but is optically thin to the cooling line emission we now
observe.  If the RSG circumstellar outflow had been denser toward the
eastern half of the equatorial plane, then the inner cavity which
the BSG wind carved out would have been closer to the central star in
that direction than toward the west.  This would have resulted in
an observed ER that is offset from the SN centroid, and an overall
density enhancement within the equatorial plane to the east.  
While we can not rule out an asymmetric explosion, values of
$v_{blast}$ suggest the shock is traveling faster into the western
half of the ER, i.e. that the asymmetry lies in the CSM.  
We thus favor the equatorial density enhancement proposed above,
which may explain both the offset ER centroid and proposed CSM
asymmetry.  This situation should have left its imprint on the
geometry of the contact discontinuity between the RSG
wind and the previously-equilibrated bubble filled by the
progenitor's main sequence wind \citep{CE89}.  This interface was
found by \citet{CKH95} at 9\arcsec--15\arcsec\ in light echoes, and
will be tested in a re-examination of those data by \citet{Sug02}.  

While the above work suggests that a fairly isotropic blast wave can
explain the appearance of most spots, it is unlikely that the blast
wave has traveled uniformly  (or continues to) in all directions into
a visibly inhomogeneous medium.  As the forward shock
approaches the ER, it must encounter an increasing density gradient,
which will cause it to decelerate on the largest scales.
This has been addressed by e.g.\
\citet{CD95,BBM97a}; and \citet{Lun99} via the low-density \ion{H}{2}
region previously mentioned.
\citet{CBM92} find that Rayleigh-Taylor
instabilities develop in the contact discontinuity between the shocked
\ion{H}{2} material and shocked ejecta.  These can develop into plumes that
redirect previously-reflected shocks back into the ER in spots
rather than uniformly \citep{BBM97b}.   We therefore expect strong
velocity inhomogeneities on small scales.  Whether or not this
mechanism is responsible for \hsvi\ and \hsvii\, both of which 
appear further from the SN than neighboring protrusions (see Figure
\ref{deproj}c), is unclear.  The probability of a randomly-placed
plume impacting an inward-facing protrusion from the ER is low,
unless the number density and/or angular size of such plumes is very
high.  This question is better addressed by hydrodynamic modeling. 

We caution the reader of the preliminary nature of these inferences.
As noted above, values of $v_{blast}$ assume an isotropic expansion at
early times, and ignore the unknown cooling time $t_c$, which roughly
dictates when a spot first becomes visible in optical line emission
after it has been shocked.  From the shock model presented above, we
see that $t_c$ is a strong function of the forward blast velocity, and
both the shape and size of the protrusion.  At this early
stage of hot spot study, all three remain highly uncertain, if not
unknown.  While our data suggest the forward blast may travel faster
in the western direction, \hsxiv, located between \hsvi\ and \hsvii,
does not require an above-average value of $v_{blast}$ to
explain its earliest appearance.  Thus, many interpretations of the
data exist.  The forward blast could be symmetric and fast ($\sim
5000$ \kms), and  \hsvi\ and \hsvii\ lie on protrusions
whose geometries yield significantly shorter cooling times (e.g. cylinders)
than those of most other spots (e.g. wide cones).  In keeping with our
proposed scenario, the geometry of \hsxiv\ could result in a much
longer cooling time than its neighboring two spots, delaying its
discovery until after \hsvi\ and \hsvii.   These ambiguous
interpretations require a combined observational and computational effort.
Continued spectroscopic monitoring of the ER and hot spots will yield
better estimates of the physical conditions of the pre- and
post-shocked gas, which serve as important constraints for numerical
modeling.   
Radiation-hydrodynamic modeling of the shock interaction with a variety of
probable protrusion geometries can offer theoretical light curves and
flux proper-motion curves against which data (such as those presented in
this paper) can be compared.  As such, the conclusions from this work
are only preliminary and suggestive, as there is much yet to unfold in
this stage of the evolution of SNR 1987A.

Finally, we note that as of 2001 April, there remain protrusions from
the ER which do not appear to host hot spots.  From Figure
\ref{deproj}, we see that hot spots are expected at rough P.A.s of 5\degr,
85\degr, 195\degr, and 310\degr.  While there is a marginal detection
of flux at the latter
position in 2001 May, no spots have been detected between
P.A. 50\degr--90\degr, and 170\degr--220\degr.  The
most recent \ion{He}{1} data show a 
smear of flux toward PA 90, suggestive that new spots may be forming
in that region, however no significant detection of flux is made
between PA 180\degr--220\degr.  The first appearance of spots in this
vicinity should prove another interesting measurement of asymmetry in
the forward blast.

%%%%%%%%%%%%%%%%%%%%%%%%%%%%%%%%%%%%%%%%%%%%%%%%%%%%%%%%%%%%%%%%%%

\subsection{ER Ellipticity and Distance to the SN \label{ssec-dist}}

The analyses of \citet{Gou98}  suggest that the
ER might have an intrinsic ellipticity with axis ratio 0.95, based in
part on the discrepancy between the observed ER geometry and that
implied by the interpretation of UV line-emission light curves as
fluorescent light echoes off the ER \citep{Pan91}.  Measuring the
inclination angle 
$i_t$ of the ER from the delay of the UV light echoes, \citet{Gou98}
find $i_t=40\fdg5 \pm 0\fdg5$.  Insertion of our geometric
inclination angle $i_{\theta}=43\fdg8 \pm 0\fdg13$ into their Equation
4.5 also yields an intrinsic ellipticity
${b}/{a}={\cos{i_{\theta}}}/{\cos{i_t}}=0.95$.  The three-dimensional
structure of the hourglass nebula containing the ER \citep{CKH95} has
been reanalyzed by A.~P.~S.~Crotts \citep[as reported in][]{Gou98},
revealing again an intrinsic flattening of the ER of 
${b}/{a}=0.95 \pm 0.02$.  The agreement of these two independent
measurements is suggestive that the ER may have an intrinsic 5\%
ellipticity.  

An ellipse, when inclined from the line of sight, is mapped to a new
ellipse, and can always be rotated about orthogonal axes into a circle
(even if that is not its actual shape).  We find however that the ER
deviates somewhat from even an
elliptical morphology.  It is unclear at this time what the actual
geometry is, however we can envision a number of possibilities.  The
ring could also be warped, as noted in \S\ref{ssec-param}, or it could
have an inherent radial profile that is not a smooth function of P.A.
Recalling the postulated density enhancement to the east, an
axially-symmetric outflow would expand furthest into the lowest
density medium.  Figure \ref{deproj} shows that the ER material to the
northwest is furthest from the SN, suggestive of this simpler
interpretation.

Assuming that the ER is planar and circular, \citet{Gou98} calculate the
distance to the SN as $D={c(t_+ + t_-)}/{\theta_+}$ where
$t_{\pm}$ are the times of the two cusps of UV line emission around
the ER following the SN explosion, and $\theta_+$ is the major
axis of the ER.  Using a re-analysis of {\em IUE} data yielding  $t_-
= 80.5 \pm 1.7$ days and $t_+=378.3 \pm
4.8$ days, and the geometric data in [\ion{O}{3}] from \citet{Pla95},
Gould \& Uza
find a weighted average distance of $47.25 \pm 0.76$ kpc.  If
intrinsic ellipticity of the ER is taken into consideration, Gould \&
Uza find $D={c(t_+ + t_-)\cos{i_{\theta}}}/{\theta_-}$, where
$\theta_-$ is the minor axis of the ER, yielding an upper limit
distance of $48.8\pm1.1$ kpc.  Finally, they consider the case of UV
emission arising from the inner edge of the ER, and find
$D<50.8\pm0.9$ kpc.  They discount this latter scenario as highly
implausible, explaining that \ion{N}{3} (seen in the UV echo) has the
same ionization potential as \ion{O}{3} (used to measure the geometric
data), making it unlikely that their physical distribution is
dissimilar.

The success of the Gould \& Uza method
relies on the assumption that ${\theta_+}$ measures the size of the
region from which the UV echo occurred, however the geometric
parameters of the ER presented in this work and \citet{Pla95} only
measure the size of the optically-emitting region at the time the data
were taken.  Both
\citet{Pla95} and \citet{SLC01b} found in {\em HST} imaging that the
ER is larger in [\ion{O}{3}] than in H$\alpha$ and [\ion{N}{2}], and
\citet{LS97} find that the [\ion{N}{2}]-emitting ER gas has a higher
density than that emitting [\ion{O}{3}].
 Recalling the proposed formation scenario in which the ER lies at the
waste of a bipolar ``hourglass'' nebula of gas, \citet{LS97} offer a
simple model in which the  [\ion{N}{2}]-emitting gas is
located within the equatorial plane, while  [\ion{O}{3}] emission
arises from less dense gas along the walls of the hourglass just above
or below this plane.  The densest gas cools the most rapidly, hence
the [\ion{N}{2}]-dominated ER seen today should have been dominated by
[\ion{O}{3}] at earlier times.  Given that the [\ion{0}{3}] ER
appeared larger than that in [\ion{N}{2}] in 1995, it is unlikely that
the values of ${\theta_+}$ from \citet{Pla95} or this work
measure the position of the dense material  which fluoresced to
produce the observed UV echo.  Rather, this component may have had
such a high density that it has faded and is ``invisible'' in optical
line-emission today. 

We propose the following pragmatic distance estimate.  The
best-fit parameters for the ER (\S\ref{ssec-param})
offer upper limits to the location of the UV-emitting region, while
the dense inner edge of the ER, now well delineated by hot spot
activity, serves as a reasonable lower limit.
Measuring the innermost edge of the ER as the average deprojected
position of all hot spots, we find
$\theta_{+,min}=1\farcs464\pm0\farcs054$, while we take
$\theta_{+,max}=1\farcs658\pm=0\farcs026$ 
from Table \ref{tbl-geo}.  Assuming the ER is planar
and circular, we now find a lower distance limit of $47.9\pm0.92$ kpc,
and an upper limit of $54.4\pm2.1$ kpc.  For comparison,
\citet{Fea99} reviews a number of distance determinations to the LMC,
finding average values of $\sim$55 kpc from Cepheids and
around 53.7 kpc from Miras and RR Lyrae, while a few methods do favor a
shorter distance of 45--48 kpc, such as that determined via eclipsing
binaries.  Using red clump stars in {\em
HST} fields surrounding the SN,
\citet{Rom00} determine the distance to the LMC to be $52.2\pm2.3$
kpc.  The light echo analyses of \citet{Xu95} show the SN to be at
least \onehalf\ kpc deep into the inclined plane of the LMC, thus the
lower distance limit appears inconsistent with these distance
indicators.  Rather, we find they favor the longer SN distance
scale, and thereby suggest that the UV echo did originate along the
inner edge of the ER, as traced by the positions of the hot spots.

%%%%%%%%%%%%%%%%%%%%%%%%%%%%%%%%%%%%%%%%%%%%%%%%%%%%%%%%%%%%%%%%%%

\section{Conclusions \label{sec-conc}}

We have shown that a hot spot is present, or appears to be
developing, on nearly every inward-protrusion or gradient of
increasing flux.  We expect to see spots developing at rough P.A.s of
5\degr, 85\degr, 195\degr, and 310\degr\ in the years 2002-2003, with the
currently confirmed spots eventually spreading in P.A.\ into resolved
emission regions.
It is {\em abundantly} clear that hot spot evolution changes
on time scales as short as one month, and dramatically within six months.
Observing campaigns, such as the ground-based IR and STIS spectral
observations from this work, are mandated to efficiently and
completely monitor this rapidly evolving system.  To properly
interpret  current and future observations of this system, models of
optical line profiles and intensities, as well as radiative shock
models consistent with the observed hot spot loci and evolution,
should be further developed.

As \citet{Mic00} point out, it will not be straightforward to develop a
quantitative model of line spectra and evolution of hot spots.
It is our hope that the data reported in this work will be combined
with that from spectra, yielding detailed information about density
and temperature \citep{Law02}, and shock velocity \citep{Mic00}, to
provide a solid empirical base on which to build appropriate analytic
models to both analyze and predict future evolution.    Nonetheless,
analytic predictions of the ER-ejecta evolution do exist.
Of particular interest, \citet{BBM97b}
find that the impact should have three periods of brightening,
corresponding to the initial impact of the blast wave, the later
impact of a reflected shock, and the merging of the two shocks within
the ring material.  Within this framework, we question whether
the ``glitch'' from \hsi\ around day 4400 (and perhaps currently
occurring in \hsvii) 
may correspond to the delay between the first two brightening
periods.  This should be accompanied by an increase in temperature,
density and shock velocity from the hot spot, all measurable through
spectroscopy.
In the meanwhile, only time will more fully ``illuminate'' the
ER-ejecta evolution.

\acknowledgements

APSC, SSL and BEKS gratefully acknowledge funding from NASA (NAG5-3502)
and STScI (GO-8806 and GO-8872).  They are also thankful for the
allocation by the director of STScI for discretionary time.  BEKS
thanks Robert Uglesich for his patient assistance with difference
imaging and systems management, Peter Lundqvist and our anonymous
referee for their thoughtful comments.

%%%%%%%%%%%%%%%%%%%%%%%%%%%%%%%%%%%%%%%%%%%%%%%%%%%%%%%%%%%%%%%%%%
%%%        figures figures figures figures figures figures     %%%
%%%%%%%%%%%%%%%%%%%%%%%%%%%%%%%%%%%%%%%%%%%%%%%%%%%%%%%%%%%%%%%%%%

\clearpage

\clearpage

%%%%%%%%%%%%%%%%%%%%%%%%%%%%%%%%%%%%%%%%%%%%%%%%%%%%%%%%%%%%%%%%%%
%%%       Tables Tables Tables Tables Tables Tables Tables     %%%
%%%%%%%%%%%%%%%%%%%%%%%%%%%%%%%%%%%%%%%%%%%%%%%%%%%%%%%%%%%%%%%%%%

\begin{deluxetable}{lcccccccc}
\rotate
\tablecaption{Hot Spots in SNR 1987A \label{tbl-pos}}
\tablewidth{0pt}
\tablehead{
 \colhead{} &
 \multicolumn{2}{c}{Observed Positions} &
\multicolumn{2}{c}{Deprojected Positions} &
 \multicolumn{3}{c}{} \\
 \colhead{Spot} & \colhead{$r (\arcsec)$} &  \colhead{${\rm P.A.} (\degr)$} &
 \colhead{$r (\arcsec)$} & \colhead{${\rm P.A.} (\degr)$} &
 \colhead{$t_{earliest}$\tablenotemark{a}} &
 \colhead{$\Delta t_{light}$\tablenotemark{b}} &
% \colhead{$\Delta t_{shock}$\tablenotemark{c}} &
 \colhead{$v_{blast}$\tablenotemark{c}} &
 \colhead{$\Delta v_{blast}$\tablenotemark{d}}
% \colhead{$v_{shock}$\tablenotemark{d}}
}
\startdata
HS1-029  & $0.560\pm 0.015$ & $ 27.8\pm   0.7$ & $0.704\pm 0.019$ & $
19.3\pm   0.8$ & $2933$ & $  143$ & $  2590\pm   940$ & 843\\
HS2-104  & $0.673\pm 0.015$ & $105.6\pm   0.7$ & $0.727\pm 0.017$ &
$113.5\pm   1.2$ & $4283$ & $  -52$ & $  3347\pm   496$ & 544\\
HS3-126  & $0.607\pm 0.016$ & $124.3\pm   0.6$ & $0.729\pm 0.020$ &
$133.6\pm   0.9$ & $4337$ & $  -99$ & $  3546\pm   582$ & 574\\
HS4-091  & $0.702\pm 0.031$ & $ 90.7\pm   1.1$ & $0.712\pm 0.032$ & $
94.6\pm   2.0$ & $4337$ & $   -3$ & $  3215\pm   903$ & 500\\
HS5-139  & $0.565\pm 0.015$ & $140.8\pm   1.2$ & $0.736\pm 0.020$ &
$148.1\pm   1.4$ & $4440$ & $ -127$ & $  3908\pm   578$ & 613\\
HS6-229  & $0.697\pm 0.034$ & $229.4\pm   1.8$ & $0.777\pm 0.039$ &
$220.7\pm   2.9$ & $4440$ & $ -131$ & $  5108\pm  1123$ & 803\\
HS7-289  & $0.707\pm 0.029$ & $290.5\pm   1.6$ & $0.785\pm 0.033$ &
$299.2\pm   2.5$ & $4440$ & $   72$ & $  4992\pm   896$ & 722\\
HS10-040 & $0.607\pm 0.034$ & $ 40.4\pm   2.3$ & $0.713\pm 0.041$ & $
31.1\pm   3.3$ & $4725$ & $  134$ & $  2747\pm  1005$ & 350\\
HS11-355 & $0.535\pm 0.035$ & $349.7\pm   2.9$ & $0.741\pm 0.049$ &
$349.9\pm   2.9$ & $4816$ & $  152$ & $  3336\pm  1148$ & 409\\
HS12-050 & $0.629\pm 0.032$ & $ 50.8\pm   2.7$ & $0.696\pm 0.037$ & $
42.2\pm   4.4$ & $4816$ & $  115$ & $  2292\pm   887$ & 285\\
HS13-165 & $0.531\pm 0.048$ & $165.5\pm   2.4$ & $0.734\pm 0.066$ &
$166.8\pm   2.5$ & $4999$ & $ -149$ & $  3281\pm  1624$ & 429\\
HS14-249 & $0.713\pm 0.036$ & $247.6\pm   2.0$ & $0.729\pm 0.037$ &
$243.0\pm   3.8$ & $4999$ & $  -79$ & $  3087\pm   891$ & 385\\
\enddata
\tablenotetext{a}{Earliest detection in {\em HST} data, days after SN.}
\tablenotetext{b}{Light-travel delay in days, measured from the SN
centroid, assuming a distance of 50 kpc.}
%\tablenotetext{c}{Expected turn-on time, in days after HS~1-029,
%based on uniform ejecta outflow at 2800 \kms.}
\tablenotetext{c}{Shock velocity, in \kms, required to travel from
$0\farcs6$ at day 1300 [the rough epoch and position at which radio
emission was first detected, see \citet{Man01}] to a spot given its
deprojected position, earliest detection and light-travel delay.
For spots 1--3, we use the early-time radial positions
shown in Figure \ref{hsmove}.}
\tablenotetext{d}{Increase in shock velocity $v_{blast}$ if a cooling
time $t_c=1$ yr is included.}  
\end{deluxetable}

\begin{deluxetable}{l c c c}
\tablecaption{Least-Squares Fits to Hot Spot Positional Data\label{tbl-mov}}
\tablewidth{0pt}
\tablehead{
\colhead{Object} & \multicolumn{3}{c}{Best-Fit Slope} \\
\colhead{} & \colhead{Observed} &
             \colhead{Projected\tablenotemark{a}} &
             \colhead{Deprojected\tablenotemark{a}} \\
\colhead{} & \colhead{mas yr$^{-1}$} & \colhead{\kms} & \colhead{\kms}
}
\startdata
\hsi   & $7.49\pm0.58$& $1781\pm138$ & $2338\pm300$ \\
\hsii  & $5.54\pm1.84$ & $1315\pm437$ & $1752\pm598$ \\
\hsiii & $10.4\pm1.5$ & $2461\pm359$ & $2966\pm774$ \\
Field Star 1 & $-0.21\pm0.39$& $-50\pm 92$ & \nodata \\
Field Star 2 & $0.008\pm0.49$ & $2\pm 116$   & \nodata \\
Field Star 3 & $-0.08\pm1.3$ & $-19 \pm 310$  & \nodata \\
\enddata
\tablenotetext{a}{Assuming a distance of 50 kpc.}
\end{deluxetable}

\begin{deluxetable}{lcccc}
\tablecaption{Geometric Parameters of the ER \label{tbl-geo}}
\tablewidth{0pt}
\tablehead{
\colhead{} & \colhead{} & \multicolumn{2}{c}{\citet{Pla95}}
 & \colhead{} \\
\colhead{Parameter} & \colhead {This Work} &
 \colhead{[\ion{N}{2}]} & \colhead{[\ion{O}{3}]} & \colhead{\citet{Bur95}}
}
\startdata
 Centroid Offset (mas E)   & $-20.00 \pm 0.69$& $-13\pm 44$ & $4\pm 11$
 & 0.\\
 Centroid Offset (mas N)   & $-1.94 \pm 0.36$  & $-31\pm 44$ & $7\pm 11$
 & 0.\\
 Axial Ratio	   	   & $0.722 \pm 0.002$& 0.711 & 0.724
 & 0.711\\
 Inclination Angle (\degr) & $43.78 \pm 0.13$ & $44 \pm 2$ & $43.7 \pm 1$
 & 44.7\\
 Semi-Major Axis (\arcsec) & $0.829 \pm 0.002$&  $0.845$ & $0.858$
 & 0.81\\
 Semi-Minor Axis (\arcsec) & $0.599 \pm 0.002$&  0.600 & 0.621
 & 0.57\\
 P.A.\ of Major Axis (\degr E of N) & $80.36\pm0.84$ & $84.1\pm3$&$88.6\pm3$
 & 81.2\\
\enddata
\end{deluxetable}

\begin{deluxetable}{lccccc|c|cc}
\rotate
\tablecaption{Detection Reliability and Measurement Accuracy \label{tbl-acc}}
\tablewidth{0pt}
\tablehead{
\colhead{Flux\tablenotemark{a} } &
\multicolumn{5}{c}{Hot Spot Detection} &
\colhead{Photometry} &
\multicolumn{2}{c}{Positional Errors (mas)} \\
\colhead {{ $10^{-16}$ \ergcms}} &
\colhead {$N_{\rm Detected}$} &
\colhead {$N_{\rm False}$} &
\colhead {$N_{\rm Missed}$} &
\colhead{\% Correct} & \colhead{\% Complete} &
\colhead{\% Error} & \colhead{{\em allstar}} & \colhead{{\em phot}}
}
\startdata
$0.42-1.26$    & 14 & 4  & 4 & 60 & 60   & \nodata& \nodata& \nodata \\
$1.26-2.53$ & 12 & 1  & 1 & 91 & 91   & 13.5 & 40.   & 53.  \\
$2.53-4.21$ & 10 & 0  & 0 & 100 & 100 & 11   & 29.   & 22. \\
$4.21-42.1$ & 24 & 0  & 0 & 100 & 100 & 6.5  & 8.2   & 17. \\
 
\enddata
\tablenotetext{a}{These flux bins correspond to $0.01-0.03$, $0.03-0.06$,
$0.06-0.1$ and $0.1-1.0$ counts sec$^{-1}$ through the F656N filter.
}
\end{deluxetable}

%%%%%%%%%%%%%%%%%%%%%%%%%%%%%%%%%%%%%%%%%%%%%%%%%%%%%%%%%%%%%%%%%%

\clearpage

\clearpage

\end{document}